\documentclass[%
 reprint,
superscriptaddress,
 amsmath,amssymb,
 aps,
prx,
floatfix,
]{revtex4-1}
\usepackage{blindtext}
\usepackage{graphicx}
\usepackage{dcolumn}
\usepackage{bm}
\usepackage[hidelinks]{hyperref}
\usepackage{array}
\newcolumntype{P}[1]{>{\centering\arraybackslash}p{#1}}
\hypersetup{
  colorlinks   = true, 
  urlcolor     = blue, 
  linkcolor    = blue, 
  citecolor   = blue 
}

\newcommand{\ket}[1]{|#1\rangle}

\def\lsim{\mathrel{\rlap{\lower4pt\hbox{\hskip1pt$\sim$}}
    \raise1pt\hbox{$<$}}}                
\def\gsim{\mathrel{\rlap{\lower4pt\hbox{\hskip1pt$\sim$}}
    \raise1pt\hbox{$>$}}}                
    
\begin{document}

\title{High-Fidelity Measurement of a Superconducting Qubit using an On-Chip Microwave Photon Counter}

\author{A. Opremcak}
\email[]{opremcak@wisc.edu}
\affiliation{Department of Physics, University of Wisconsin-Madison, Madison, WI 53706, USA}
\author{C. H. Liu}
\affiliation{Department of Physics, University of Wisconsin-Madison, Madison, WI 53706, USA}
\author{C. Wilen}
\affiliation{Department of Physics, University of Wisconsin-Madison, Madison, WI 53706, USA}
\author{K. Okubo}
\affiliation{Department of Physics, University of Wisconsin-Madison, Madison, WI 53706, USA}
\author{B. G. Christensen}
\altaffiliation[Present address: ]{Northrop Grumman Corporation, Linthicum, Maryland 21090, USA}
\affiliation{Department of Physics, University of Wisconsin-Madison, Madison, WI 53706, USA}

\author{D. Sank}
\affiliation{Google Research, Santa Barbara, CA 93117,  USA}
\author{T. C. White}
\affiliation{Google Research, Santa Barbara, CA 93117,  USA}
\author{A. Vainsencher}
\affiliation{Google Research, Santa Barbara, CA 93117,  USA}
\author{M. Giustina}
\affiliation{Google Research, Santa Barbara, CA 93117,  USA}
\author{A. Megrant}
\affiliation{Google Research, Santa Barbara, CA 93117,  USA}
\author{B. Burkett}
\affiliation{Google Research, Santa Barbara, CA 93117,  USA}
\author{B. L. T. Plourde}
\affiliation{Department of Physics, Syracuse University, Syracuse, NY 13244, USA}
\author{R. McDermott}
\email[]{rfmcdermott@wisc.edu}
\affiliation{Department of Physics, University of Wisconsin-Madison, Madison, WI 53706, USA}

\date{\today}

\begin{abstract}
We describe an approach to the high-fidelity measurement of a superconducting qubit using an on-chip microwave photon counter. The protocol relies on the transient response of a dispersively coupled measurement resonator to map the state of the qubit to ``bright" and ``dark" cavity pointer states that are characterized by a large differential photon occupation. Following this mapping, we photodetect the resonator using the Josephson Photomultipler (JPM), which transitions between classically distinguishable flux states when cavity photon occupation exceeds a certain threshold. Our technique provides access to the binary outcome of projective quantum measurement at the millikelvin stage without the need for quantum-limited preamplification and thresholding at room temperature. We achieve raw single-shot measurement fidelity in excess of 98\% across multiple samples using this approach in total measurement times under 500~ns. In addition, we show that the backaction and crosstalk associated with our measurement protocol can be mitigated by exploiting the intrinsic damping of the JPM itself. 
\end{abstract}

\maketitle
\section{Introduction\label{sec:level1}}

\begin{figure}[t!]
    \includegraphics[]{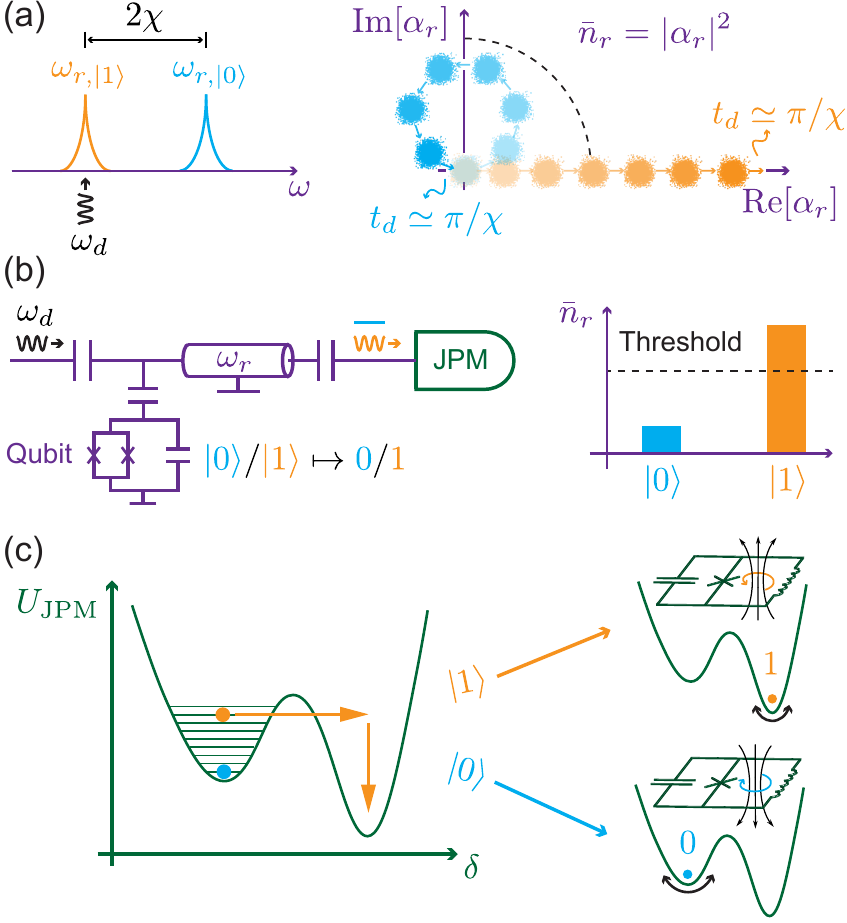}
     \caption{Qubit measurement with a photon counter. \textbf{(a)} Dispersive coupling of the qubit and the readout resonator yields distinct dressed frequencies of the cavity $\omega_{r, \ket{0}}$ and $\omega_{r, \ket{1}}$ corresponding to qubit states $\ket{0}$ and $\ket{1}$, respectively. The application of a microwave drive at frequency $\omega_d \simeq \omega_{r, \ket{1}}$ displaces the 
     photon field inside the resonator in a qubit state-dependent manner. For resonant drive (orange), the field displaces along a single quadrature axis (drawn as $\text{Re}[\alpha_r]$, where $\ket{\alpha_r}$ is the coherent state of the resonator). For off-resonant drive (blue), the readout cavity acquires a transient occupation; however, the cavity state coherently oscillates back toward vacuum at a time $\pi/\chi$, where $2\chi = \omega_{r, \ket{0}}-\omega_{r, \ket{1}}$. Therefore, drive for duration $t_d = \pi/\chi$ maps the qubit state to ``bright" and ``dark" cavity pointer states with large differential photon occupation. \textbf{(b)} Following pointer state preparation, we photodetect the resonator using the JPM, which acts as a threshold discrimiantor of microwave photon occupation $\bar{n}_r$. \textbf{(c)} Resonant interaction of the JPM with the cavity leads to conditional excitation of the JPM followed by a tunneling transition between classically distinguishable flux states of the device.
\label{fig:fig_1}}
\end{figure}

Fast, accurate state measurement is critical to the implementation of quantum error correction \cite{Fowler2012}, and global optimization of a large-scale quantum processor demands minimization of physical resources required for qubit measurement \cite{McDermott2018}. 
Prior work on the measurement of superconducting qubits has focused on suppression of errors through combined improvements in measurement speed \cite{Jeffrey2014, Walter2017, Heinsoo2018} and near-quantum-limited preamplification of the measurement signal \cite{Mutus2014, Macklin2015}; however, the physical footprint of the superconducting amplifiers, nonreciprocal circuit elements, and heterodyne detectors required to implement high-fidelity amplifier-based qubit measurement represents a significant obstacle to scaling. There have been efforts to minimize the hardware overhead associated with qubit measurement using Josephson circulators and directional amplifiers \cite{Sliwa2015, Lecocq2017, Chapman2017, Thorbeck2017, Abdo2019}, but the instantaneous bandwidths of these elements are at present too small to support multiplexed qubit measurement, the primary advantage of amplifier-based approaches \cite{Jeffrey2014, Heinsoo2018}. In related work, state-of-the-art measurement efficiencies were achieved by directly embedding a qubit within a Josephson parametric amplifier \cite{Eddins2019}; however, 
this approach is not amenable to integration with large-scale multiqubit arrays. While continued research in these directions is certainly essential, it is clear that there are major obstacles to be overcome.

In this work, we pursue an alternative approach to the measurement of  superconducting qubits based on integrated microwave photon counters. The measurement protocol relies on the transient response of a dispersively coupled linear resonator to map the state of the qubit onto ``bright" and ``dark" cavity pointer states characterized by a large differential photon occupation \cite{Blais2004, Govia2014} [Fig. \ref{fig:fig_1}{(a)}]. Following this mapping, we photodetect the resonator using the Josephson Photomultipler (JPM) \cite{Chen2011, Opremcak2018}, which operates as a threshold detector of microwave photon occupation 
[Fig. \ref{fig:fig_1}{(b)}]. The JPM is based on a capacitively shunted rf Superconducting QUantum Interference Device (SQUID) with circuit parameters chosen to yield a double-well potential energy landscape \cite{Barone1982}. JPM photodetection involves resonant transfer of energy from the bright pointer state of the readout cavity to the JPM mode, followed by a tunneling transition that changes the flux state of the JPM [Fig. \ref{fig:fig_1}{(c)}]; when the readout cavity is prepared in the dark state, no tunneling transition occurs. The flux state of the JPM represents a classical bit -- the outcome of projective quantum measurement -- that in principle can be accessed at the millikelvin stage, without the need for heterodyne detection and thresholding at room temperature. Without any fine tuning of qubit or JPM parameters, we achieve raw single shot measurement fidelities (uncorrected for qubit relaxation and initialization errors) in excess of 98\% for total measurement times around 500~ns. While the current experiments involve chips comprising two qubits, each with its own dedicated JPM, the approach can be scaled to arbitrary system size, as the physical footprint of the JPM is well matched to the footprint of the qubit. JPM-based measurement requires at most one additional flux bias line per qubit channel, while greatly relaxing the physical resources needed downstream of the millikelvin stage.   


This manuscript is organized as follows. In Section \ref{sec:level2}, we discuss the design and characterization of our qubit-JPM circuit and provide a detailed description of the qubit measurement sequence. In Section \ref{sec:level3}, we describe optimization of photon number contrast of the cavity pointer states with respect to resonator drive parameters. In Section \ref{sec:level4}, we analyze the performance of the JPM-based measurement protocol and present a detailed fidelity budget.
In addition, we discuss the long-term stability of the measurement and demonstrate the robustness of our protocol with respect to device-to-device variation. In Section \ref{sec:level5}, we discuss backaction and measurement crosstalk, and we demonstrate that intrinsic damping of the JPM itself is a resource that can be exploited to suppress initialization and crosstalk errors. In addition, we explore the degradation of measurement fidelity as the measurement cycle time is pushed below 10~$\mu$s. Finally, in Section \ref{sec:level6}, we conclude and discuss prospects for the construction of a scalable quantum-to-classical interface at millikelvin temperatures.


\section{Circuit Design and Bring-up\label{sec:level2}} 

\begin{figure}[t!]
    \includegraphics[scale=1.0]{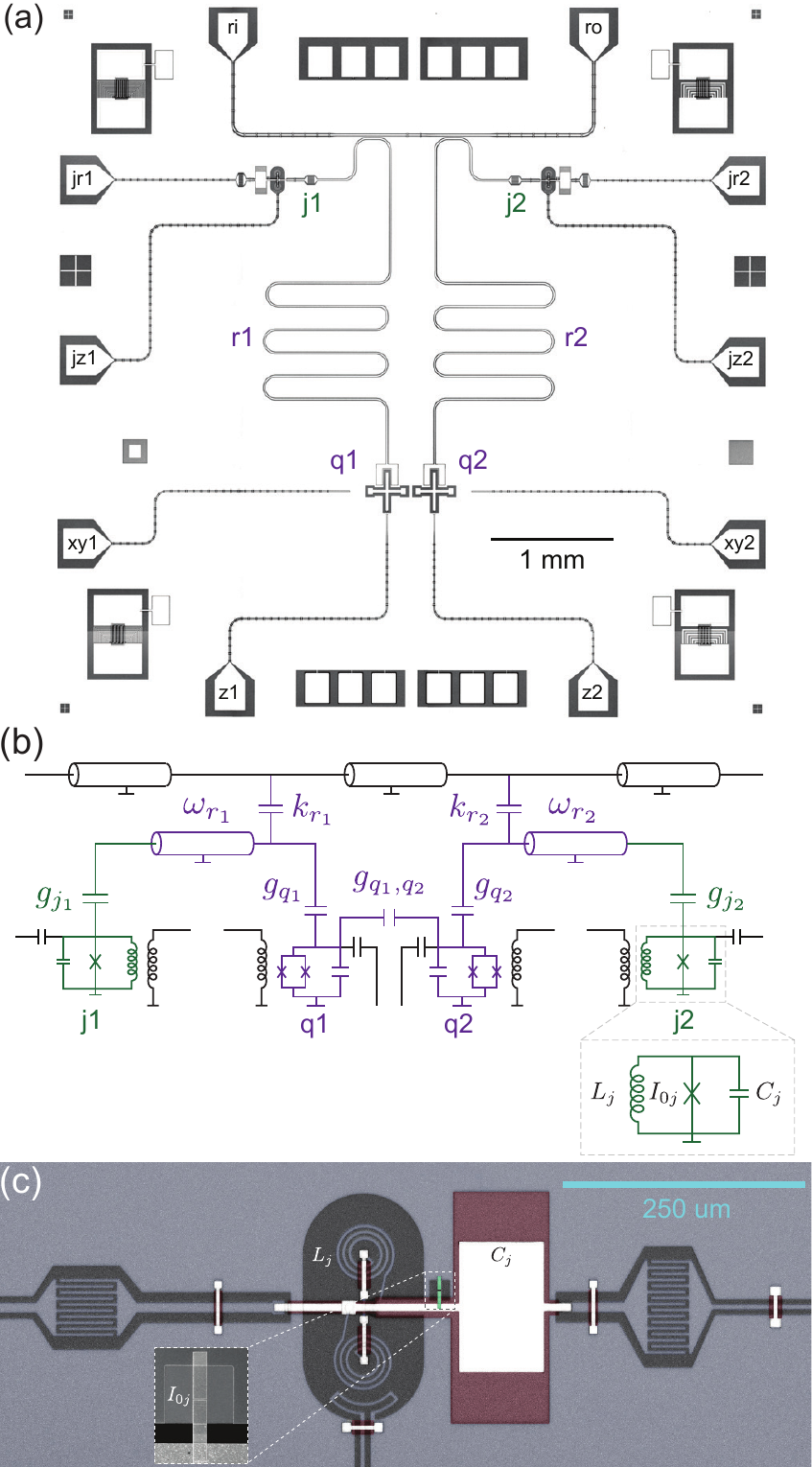}
     \caption{Device layout. \textbf{(a)} Optical micrograph of the circuit with overlaid text to indicate port functionality and the locations of critical circuit components. Each qubit-JPM system incorporates a transmon qubit q1(2), with excitation line xy1(2) and flux bias line z1(2), and a JPM j1(2), with dedicated readout line jr1(2) and flux bias line jz1(2). Each qubit-JPM pair is coupled to a half-wave CPW resonator r1(2). \textbf{(b)} Schematic diagram of the circuit. \textbf{(c)} False-color micrograph of the JPM element. \label{fig:fig_2}}
\end{figure}

Our circuit consists of two coupled qubit-JPM systems integrated onto a single silicon chip as shown in the micrograph of Fig. \ref{fig:fig_2}{(a)}. The circuit schematic is shown in Fig. \ref{fig:fig_2}{(b)}, which introduces notation that will be used throughout the text. In this section, we report the parameters of qubit-JPM pair q1-j1 on chip \#1; parameters for the other qubit-JPM pairs can be found in Table \ref{fig:tab4}. For information about sample fabrication and control wiring, see Appendices \ref{sec:sample_fabrication} and \ref{sec:experimental_setup}, respectively.

 \begin{figure}[t!]
    \includegraphics[]{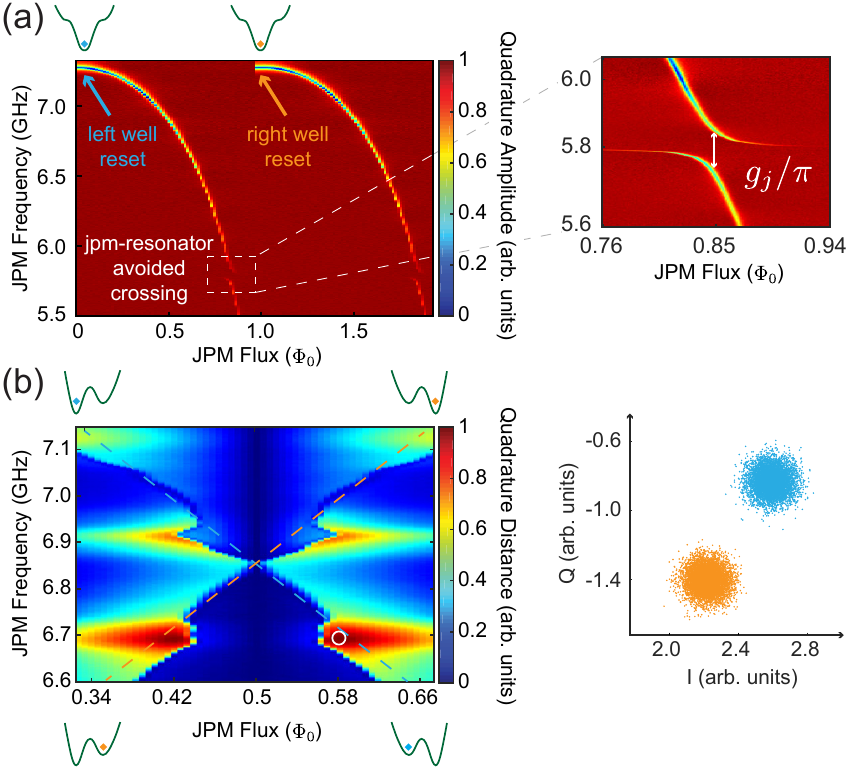}
     \caption{JPM bring-up. \textbf{(a)} JPM spectroscopy versus external flux. The spectroscopy signal is acquired from a reflection measurement at JPM readout ports jr1(2). Heterodyne detection of this signal yields in-phase (I) and quadrature (Q) components that depend on the applied frequency, the external flux bias of the JPM, and the flux state of the device. Arrows indicate left- and right-well reset bias points, where the potential energy landscape of the JPM supports only a single minimum. The enlargement to the right shows the JPM-resonator avoided level crossing. Following pointer state preparation, the JPM is biased to this point to induce resonant excitation of the JPM by the bright pointer. \textbf{(b)} Contrast in IQ signal for reflection from the JPM prepared in the left and right wells of the double-well potential. The white circle indicates the optimal parameters for JPM readout. IQ clouds for JPM readout at this point are shown on the right; here, the separation fidelity is better than 99.99\% for a readout time of 250 ns.\label{fig:fig_3}}
\end{figure}



The qubit-JPM system incorporates a frequency-tunable transmon that is dispersively coupled to a half-wave coplanar waveguide (CPW) measurement resonator \cite{Wallraff2005, Koch2007, Barends2013} with bare frequency $\omega_r/2\pi = 5.693~\text{GHz}$ and qubit-resonator coupling strength $g_{q,r}/2\pi =90~\text{MHz}$. The total energy decay rate of the measurement resonator $\kappa_r = 1/(1.53~\mu\text{s})$, which is approximately two orders of magnitude smaller than that for a typical Purcell-filtered design \cite{Jeffrey2014, Walter2017, Heinsoo2018}. The transmon has a maximum transition frequency $\omega_q/2\pi = 5.95~\text{GHz}$ and an anharmonicity $\eta/2\pi = -225~\text{MHz}$. To avoid Purcell suppression of the qubit energy relaxation time \cite{Houck2008}, we operate at qubit frequencies below $5.1~\text{GHz}$, which corresponds to a Purcell limit to qubit $T_1$ of $66~\mu\text{s}$. We remark on a distinct advantage of our approach to qubit measurement compared to amplifier-based implementations: by reading out the measurement resonator with the JPM, we avoid the usual tradeoff between measurement speed and Purcell limit to $T_1$, as coupling of the measurement resonator to its readout environment can be tuned over a broad range on nanosecond timescales by appropriate variation of the JPM bias point. In principle, the value of $\kappa_r$ can be made arbitrarily small without affecting the measurement speed; as a practical matter, however, a balance must be struck to ensure that the power delivered to the measurement resonator is sufficient for creation of the bright pointer state.

At the opposite voltage antinode, the measurement resonator is capacitively coupled to the JPM with coupling strength $g_{j,r}/2\pi = 62~\text{MHz}$. This coupling strength is optimal, as it corresponds to a half-swap period $\pi/(2 g_{j,r}) = 4~\text{ns}$ and is thus compatible with GS/s waveform generation and comparable to the energy relaxation time of the JPM $T_{1,j} = 5~\text{ns}$. The JPM circuit is formed by the parallel combination of 
a 3+3-turn gradiometric loop with inductance $L_j=1.3~\text{nH}$, a parallel-plate capacitance $C_j=2.2~\text{pF}$, and a single Al-AlO$_\text{x}$-Al Josephson junction with critical current $I_{0j}=1.4~\mu\text{A}$ [see Fig. \ref{fig:fig_2}{(b, c)}]. The plasma frequency of the JPM is tunable with external flux from 4 to 7.3~GHz, allowing for both resonant and far-detuned interactions with the measurement resonator. To retrieve qubit measurement results from the JPM, the circuit is read out in reflection using the capacitively coupled readout port labeled jr1(2) in Fig. \ref{fig:fig_2}{(a)}. The two metastable flux states of the JPM correspond to distinct plasma frequencies; therefore, microwave reflectometry in the vicinity of these resonances encodes the JPM flux onto the amplitude and phase of the reflected signal.


Device bring-up begins with JPM spectroscopy versus external flux, which yields the locations of the reset bias points that initialize the JPM in the left and right wells of its double-well potential
along with the JPM-resonator avoided level crossing [see Fig. \ref{fig:fig_3}{(a)}]. Next, we maximize contrast of JPM reflectometry for states prepared in the left and right wells over the space of JPM readout flux, measurement frequency, and JPM drive power [Fig. \ref{fig:fig_3}{(b)}]. Using optimized parameters, the fidelity with which we read out the flux state of the JPM is better than 99.99\%. In the following, we always initialize the JPM in the left well of its potential and refer to the probability of a transition to the right well as the \textit{tunneling probability}.

 \begin{figure*}[t!]
    \includegraphics[]{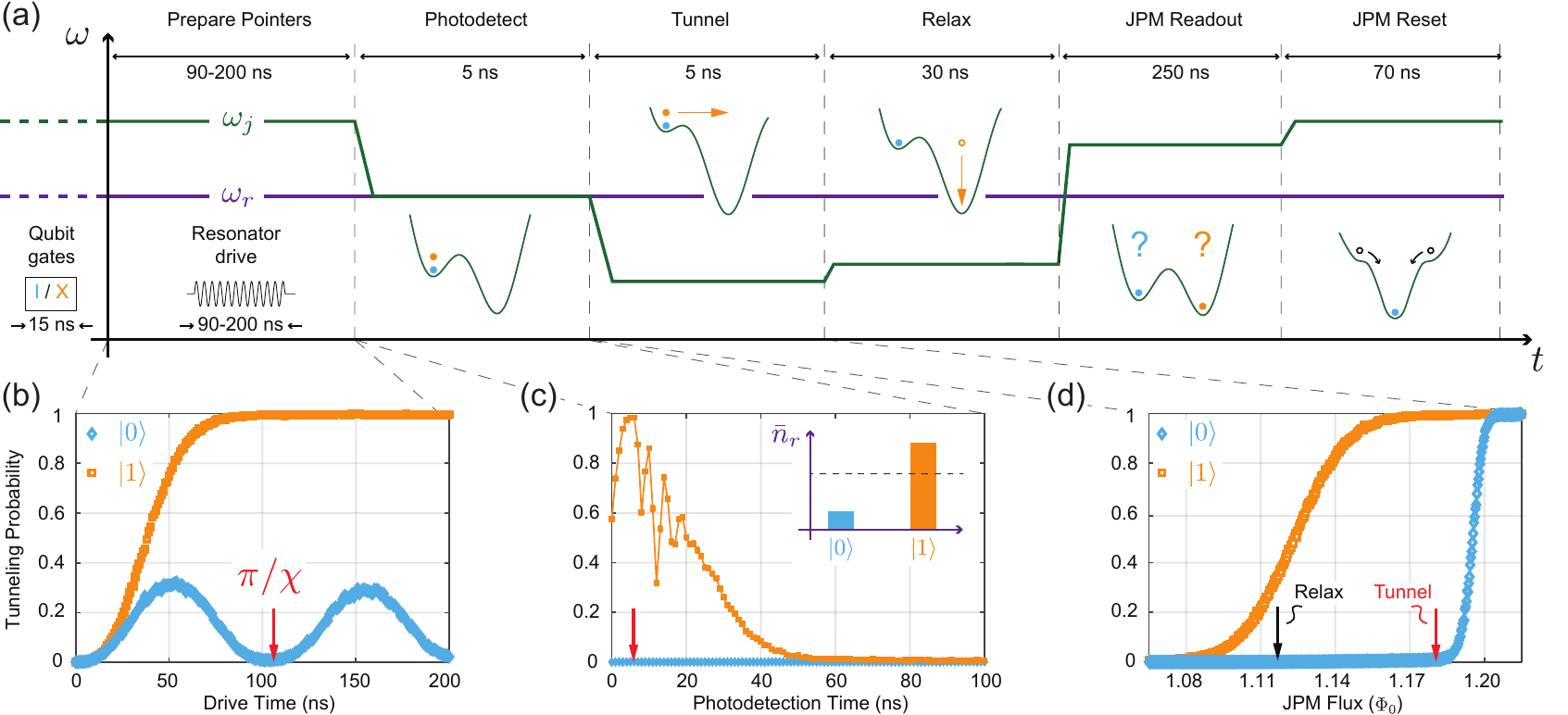}
     \caption{JPM-based qubit measurement sequence. \textbf{(a)} Measurement timing diagram; see text for detailed discussion. \textbf{(b)} Time evolution of high-contrast microwave cavity pointer states as detected by the JPM for qubits initialized in states $\ket{0}$ (blue) and $\ket{1}$ (orange). The optimal time for pointer state drive is indicated by the red arrow. \textbf{(c)} JPM tunneling probability versus photodetection time for qubits prepared in states $\ket{0}$ and $\ket{1}$. The optimal time for JPM-cavity interaction is indicated by the red arrow. \textbf{(d)} JPM tunneling probability versus tunnel bias amplitude for qubits prepared in states $\ket{0}$ and $\ket{1}$. The S-curves are well separated, corresponding to a raw measurement fidelity of $98.4$\% (see Section \ref{sec:level4}). The optimal tunnel bias point is indicated by the red arrow. Following the tunneling step, the JPM bias point is adjusted to the location indicated by the black arrow to allow the tunneled phase particle to relax. Following the tunneling event, the flux state of the JPM is read out in reflection using the methods discussed in Fig. \ref{fig:fig_3}{(b)}. Finally, the JPM is reset into the left-well state. \label{fig:fig_4}}
\end{figure*}

A timing diagram of the qubit measurement sequence is shown in Fig. \ref{fig:fig_4}{(a)}; the cartoon insets depict the evolution of the JPM phase particle during critical steps of the measurement sequence. The duration of each step is indicated at the top of each panel; for clarity, the time axis is not drawn to scale. During qubit operations prior to measurement, the JPM is biased at its flux-insensitive upper sweet spot to minimize JPM-induced damping of the measurement resonator. We prepare the target qubit state by applying the $X$-gate ($I$-gate); to achieve high-fidelity single-qubit gates, we implement fast (15~ns-long) cosine-shaped derivative reduction by adiabatic gate (DRAG) pulses with a static detuning correction \cite{Motzoi2009, Lucero2010, Chen2016}.  At the start of the measurement sequence, microwave drive at frequency $\omega_{r, \ket{1}}$ is used to prepare the bright (dark) pointer state. In Fig. \ref{fig:fig_4}{(b)}, we show the time evolution of optimized pointer states as detected by the JPM  (see Section \ref{sec:level3} for methods); the resonator drive time $t_d=105~\text{ns}$ for the datasets shown in Fig. \ref{fig:fig_4}{(c, d)}. Next, the JPM is biased into resonance with the measurement resonator to induce intrawell excitations of the phase particle conditioned on the qubit state \cite{Hofheinz2008}. The energy transferred into the JPM is maximal for a photodetection time of $5~\text{ns}\approx \pi/2 g_{j,r}$ [Fig. \ref{fig:fig_4}{(c)}]. This timescale is independent of the photon occupation in the resonator, as one expects for coupled harmonic systems: at the JPM-resonator avoided level crossing, the left well of the JPM supports approximately 50 bound states. 
Immediately following photodetection, the JPM is biased towards the critical flux at which the shallow minimum in the potential energy landscape vanishes in order to induce interwell tunneling of excited states [Fig. \ref{fig:fig_4}{(d)}] \cite{Cooper2004}. The duration and amplitude of this bias pulse are chosen to maximize tunneling contrast between qubit excited and ground states; the optimal tunnel bias point is indicated by the red arrow in Fig. \ref{fig:fig_4}{(d)}. We then adjust the JPM bias to the location indicated by the black arrow in Fig. \ref{fig:fig_4}{(d)} to allow the tunneled phase particle to relax for 30~ns. Without this step, a small fraction ($\sim5\%$) of the tunneled population migrates back into the left well, resulting in a degradation of measurement fidelity. To retrieve the qubit measurement results, we read out the JPM state using the methods discussed in Fig. \ref{fig:fig_3}{(b)}. Finally, the JPM is reset into the left-well state for use in subsequent experiments. 

\section{Pointer State Preparation\label{sec:level3}}
\begin{figure*}[t!]
    \includegraphics[]{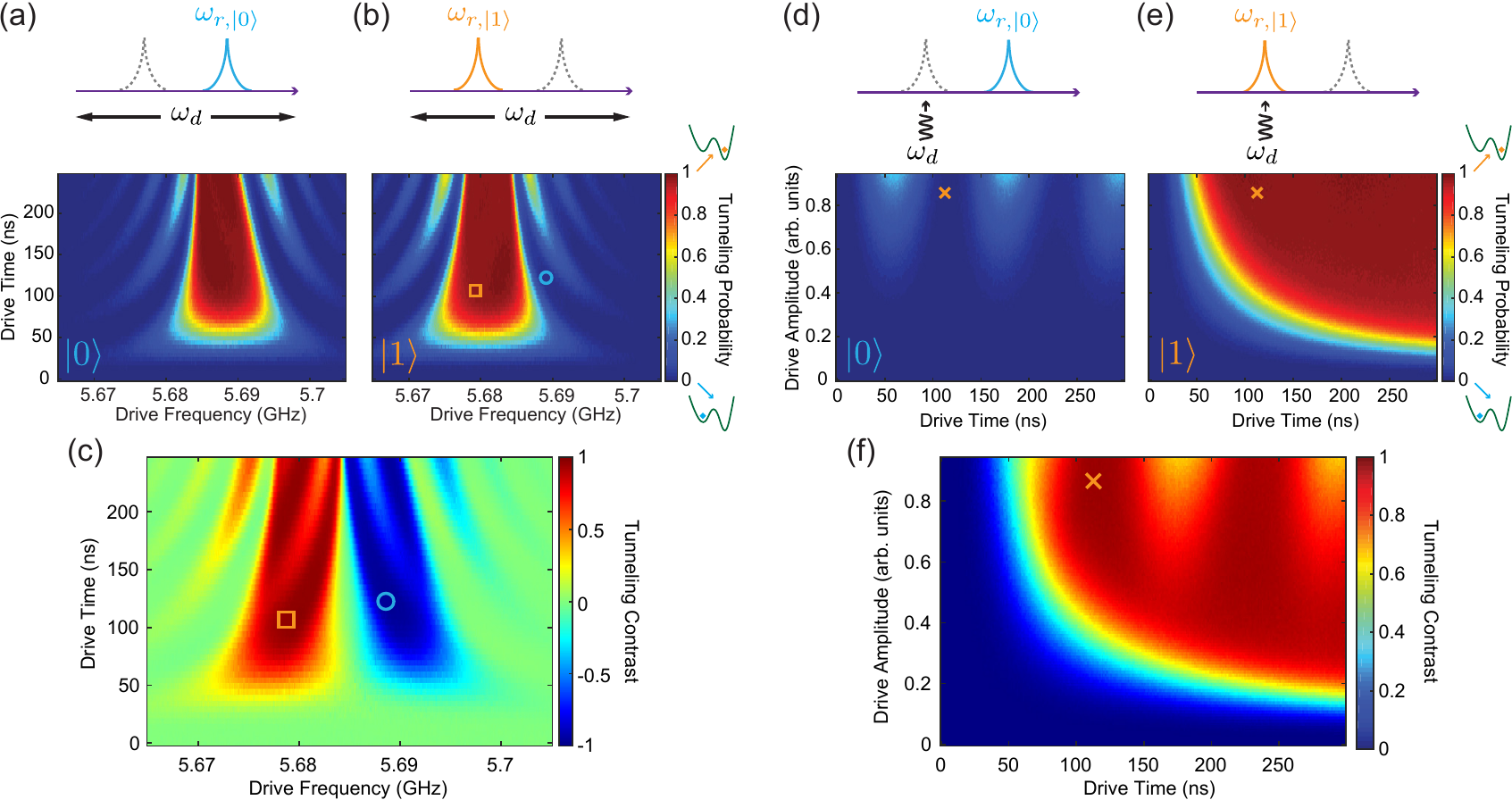}
     \caption{Pointer state preparation. \textbf{(a)} JPM tunneling probability versus resonator drive time and frequency with the qubit prepared in $\ket{0}$. \textbf{(b)} As in (a), but with the qubit prepared in $\ket{1}$. \textbf{(c)} The difference in these scans allows determination of the optimal drive frequency and time that maximize single-shot measurement fidelity. We find two local maxima in measurement fidelity for drive frequencies near $\omega_{r, \ket{0}}$ (overlaid circles) and $\omega_{r, \ket{1}}$ (overlaid squares) for a duration $t_d \simeq \pi/\chi$. For comparison with parts (d)-(e), these datasets were taken at a resonator drive amplitude of 0.8 arb. units. \textbf{(d)} JPM tunneling probability versus resonator drive amplitude and time with the qubit prepared in the $\ket{0}$ state. This scan uses the drive frequency $\omega_d \simeq \omega_{r, \ket{1}}$ found in parts (a) and (b). \textbf{(e)} As in (d), but with the qubit prepared in $\ket{1}$. \textbf{(f)} The difference in these scans yields the optimal drive amplitude and time for pointer state preparation as indicated by the overlaid X symbols. \label{fig:fig_5}}
\end{figure*}

The success of our measurement protocol hinges on our ability to create high-contrast microwave cavity pointer states conditioned on the state of the qubit [see Fig. \ref{fig:fig_1}]. To achieve this experimentally, we need to determine the optimal resonator drive frequency, time, and amplitude. To optimize pointer state preparation, we begin with two-dimensional scans of the resonator with sweeps of both drive frequency and time, as shown in Fig. \ref{fig:fig_5}{(a, b)}. Both datasets are taken over identical ranges and differ only in the prepared qubit state. The cartoons above each plot indicate that we are scanning over a range of frequencies containing both dressed resonances of the cavity, with the dressed resonance corresponding to the prepared qubit state drawn using a solid line. Optimal measurement contrast is achieved at drive parameters that maximize the difference in tunneling probability for the prepared qubit states $\ket{0}$ and $\ket{1}$ [Fig. \ref{fig:fig_5}{(c)}]; the optimal parameters correspond roughly to cavity drive at frequency $\omega_d = \omega_{r, \ket{1}}$ (overlaid squares) and $\omega_d = \omega_{r, \ket{0}}$ (overlaid circles) for a duration $t_d \simeq \pi/\chi$. Slight deviation of the optimal drive frequency from the two dressed cavity resonance frequencies and of the optimal drive time from $\pi/\chi$ can be understood as the result of nonlinearity of the measurement resonator inherited from the qubit; this nonlinearity similarly limits the size of the bright pointer state that can be created with a naive cavity ringup pulse applied at fixed frequency. As the dressed cavity resonance corresponding to qubit $\ket{1}$ disperses less strongly with power than the resonance corresponding to qubit $\ket{0}$, we achieve best measurement fidelity with cavity drive $\omega_d \simeq \omega_{r, \ket{1}}$, meaning that the qubit $\ket{1}$ ($\ket{0}$) state is mapped onto the bright (dark) cavity pointer state. 

Next, we perform two-dimensional scans of the resonator with sweeps of both drive amplitude and time, as shown in Fig. \ref{fig:fig_5}{(d, e)}. The cartoons above each plot indicate the frequency of the cavity drive with respect to the dressed cavity resonances. \textcolor{black}{Taking the difference between these scans yields the optimal drive amplitude and time as shown in Fig. \ref{fig:fig_5}{(f)}}. Scans of type Fig. \ref{fig:fig_5}{(a, b)} and Fig. \ref{fig:fig_5}{(d, e)} are repeated iteratively to optimize single-shot measurement fidelity over the space of resonator drive time, frequency, and amplitude, with the final results displayed in Fig. \ref{fig:fig_4}{(b)}. This method converges on a drive frequency that is $-2.1~\text{MHz}$ detuned from $\omega_{r, \ket{1}}/2\pi$, leading to a $22\%$ decrease in the resonator drive time as compared to $\pi/\chi$. The bright pointer state corresponds to a mean resonator occupation $\bar{n}_r \approx 27~\text{photons}$, determined via the ac Stark effect (see Appendix \ref{sec:stark_calibration} for further detail). Ultimately, photon number contrast is limited by imperfect preparation of the dark pointer state: as occupation of the dark pointer becomes comparable to the critical photon number $n_\text{crit} = (\Delta_{q,r}/g_{q,r})^2/4$ \cite{Blais2004}, the nonlinearity of the resonator prevents coherent oscillation back to the vacuum state \cite{Khezri2016, Khezri2017}, contributing an infidelity around 0.6\% (see discussion in the next section). We expect that it will be straightforward to suppress this source of infidelity by a slightly more complicated ringup sequence involving either composite pulses or a chirped frequency drive.

\section{Measurement Fidelity\label{sec:level4}}
\begin{figure}[t!]
    \includegraphics[scale=1.0]{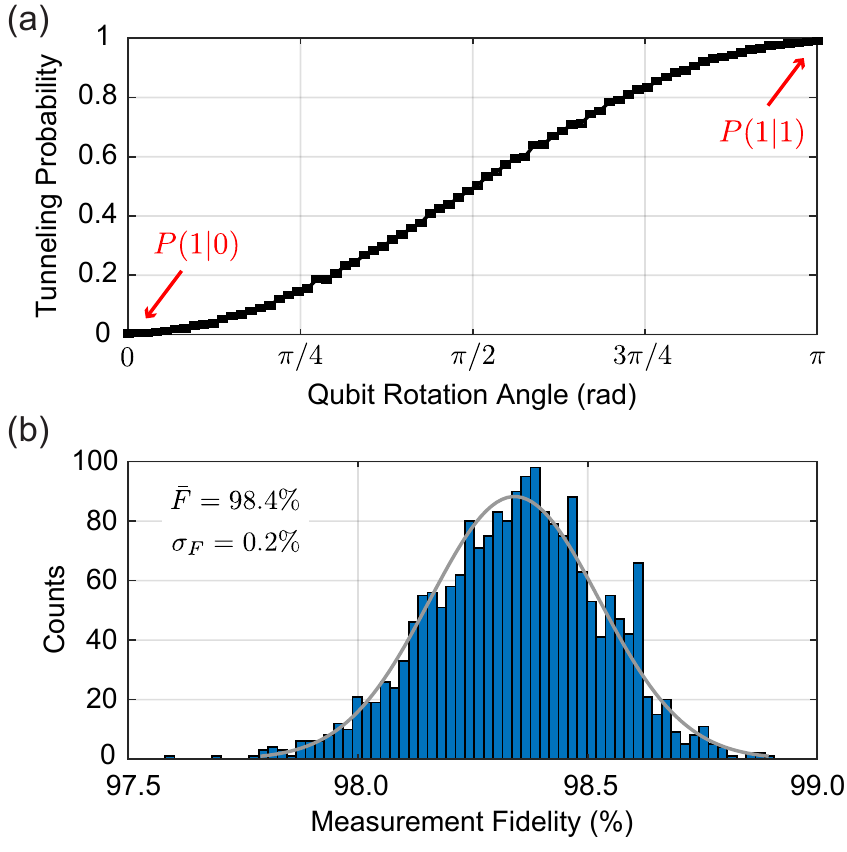}
     \caption{Measurement fidelity and long-term stability. \textbf{(a)} JPM tunneling probability versus qubit rotation angle to identify values for $P(1|0)$ (0 rotation) and $P(1|1)$ ($\pi$ rotation). For this dataset, $P(1|0) = 0.4\%$ and $P(1|0) = 99.0\%$. \textbf{(b)} Histogram of measurement fidelity $F$ logged over the span of twelve hours (20,000 independent measurements of $F$), demonstrating the robustness of JPM-based measurement with respect to long-term drift. A Gaussian fit to the histogram yields an average fidelity $\bar{F}=98.4\%$ with standard deviation $\sigma_F=0.2\%$. \label{fig:fig_6}}
\end{figure}

We analyze the performance of JPM-based measurement in terms of the fidelity
\begin{equation}
F = 1 - P(1|0)-P(0|1),
\end{equation}
where $P(i|j)$ is the probability of measuring the qubit in state $\ket{i}$ given that it was nominally prepared in state $\ket{j}$ \cite{Gambetta2007}; here, detection of a tunneling transition from the left well to the right well of the JPM constitutes measurement of the qubit $\ket{1}$ state, while the absence of a tunneling transition constitutes measurement of qubit $\ket{0}$. Using the measurement sequence described in Fig. \ref{fig:fig_4}, we perform a standard Rabi experiment to identify values for $P(1|1)$ and $P(1|0)$ as shown in Fig. \ref{fig:fig_6}(a); assuming that leakage errors are negligible, we have $F=P(1|1)-P(1|0)$. In order to faithfully estimate the conditional probabilities $P(i|j)$, the measurement sequence is repeated 5,000 times. Prior to each measurement, an active qubit reset step is performed to extract unwanted excess $\ket{1}$ population from the qubit (see Appendix \ref{sec:resonator_qubit_reset} for further detail); without this step, the excess $\ket{1}$ state population of our qubits is approximately $4\%$.

To characterize the long-term stability of JPM-based measurement, we perform 20,000 independent determinations of $F$ evenly spaced over the span of twelve hours; the results are shown in the histogram of Fig. \ref{fig:fig_6}(b). We achieve an average raw measurement fidelity $\bar{F}=98.4\pm0.2\%$, uncorrected for state preparation, relaxation, or gate errors. A detailed budget of measurement infidelity is shown in Table \ref{tab:tab1}. The nonvanishing $P(1|0)$ contains contributions both from qubit initialization errors and from imperfect dark pointer state preparation. Using the methods described in Appendix \ref{sec:excess_one_state}, we infer an excess $\ket{1}$ population of 0.3\% following active qubit reset. This initialization infidelity degrades both $P(0|0)$ and $P(1|1)$, contributing an overall infidelity of 0.6\% to our measurement. We attribute the remaining portion of $P(1|0)$ to imperfect dark pointer state preparation, for which we obtain 0.6\%. Nonvanishing $P(0|1)$ contains additional contributions from qubit relaxation and $X$-gate error. Qubit relaxation with timescale $T_1$ = 16.9~$\mu$s contributes an infidelity $t_d/2T_1$ = 0.3\%, where $t_d$ = 105~ns is the drive time for pointer state preparation. Finally, we use interleaved randomized benchmarking \textcolor{black}{(IRB)} \cite{Magesan2012} to characterize the infidelity of our $X$-gate, for which we find 0.1\%. 

We have characterized measurement fidelity for system q1-j1 on chip \#1 over a range of qubit operating points, corresponding to a range of optimal resonator drive times from 90-200~ns; results are shown in rows 1-4 of Table \ref{tab:tab2}. For all experiments, we maintain the same readout parameters calibrated at the initial qubit bringup point $\omega_q/2\pi=5.037$~GHz, apart from the resonator drive frequency and the resonator drive time, which must be matched to $\pi/\chi$. We maintain similar performance across all four qubit frequencies. This demonstrates that fine-tuning of JPM bias parameters is not needed to address qubits that resonate over a broad range of frequencies.

While the above results were obtained for the single qubit-JPM pair q1-j1 on chip \#1, we observe similar performance for the three other qubit-JPM pairs that we have examined; measurement fidelities for these devices are reported in rows 5-7 of Table \ref{tab:tab2}. The durations of the flux bias parameters determined from our bring-up of pair q1-j1 on chip \#1 were used for all remaining qubit-JPM pairs, without full optimization of each separate qubit-JPM system. The raw single-shot measurement fidelity averaged over the four qubit-JPM pairs is 98\%. 



\begin{table}[h]
\begin{tabular}{ |P{3.3cm}|P{1.5cm}|P{3cm}| }
 \hline
 Source of Infidelity & Infidelity & Calculation Method\\\hline
 Excess $\ket{1}$ population  & 0.6\%  & low power drive\\
 Imperfect dark pointer  & 0.6\%  & high power drive\\
 Qubit relaxation  & 0.3\% & $t_d/2 T_1$\\
 $X$-gate  & 0.1\%  & IRB \cite{Magesan2012}\\
 \hline
\end{tabular}
\caption{Infidelity budget for the data displayed in Fig. \ref{fig:fig_6}{(b)}. \label{tab:tab1}}
\end{table}

\begin{table}[h]
\begin{tabular}{ |P{0.75cm}|P{1.75 cm}|P{1.25cm}|P{1.75cm}|P{2cm}| }
 \hline
 Chip \# & Qubit-JPM Pair & $\omega_q/2\pi$ (GHz) & Resonator Drive Time & Measurement Fidelity\\
\hline
 1  & q1-j1 & 5.037 & 105 ns & 98.4\%\\
 1  & q1-j1 & 5.098 & 90 ns & 98.3\%\\
 1  & q1-j1 & 4.980 & 150 ns & 97.1\%\\
 1  & q1-j1 & 4.833 & 200 ns & 98.1\% \\
 1  & q2-j2 & 5.069 & 147 ns & 98.0\%\\
 2  & q1-j1 & 5.068 & 128 ns & 97.6\%\\
 2  & q2-j2 & 5.062 & 163 ns & 98.3\%\\
 \hline
\end{tabular}
\caption{Measurement fidelity within and across devices. The first entry corresponds to the data shown in Fig. \ref{fig:fig_6}{(b)}, and therefore represents the average fidelity $\bar{F}$. The remaining entries (rows 2-7) correspond to single measurements of $F$. \label{tab:tab2} }
\end{table}

\section{Backaction and Crosstalk\label{sec:level5}}
\begin{figure}[t!]
    \includegraphics[scale=1.0]{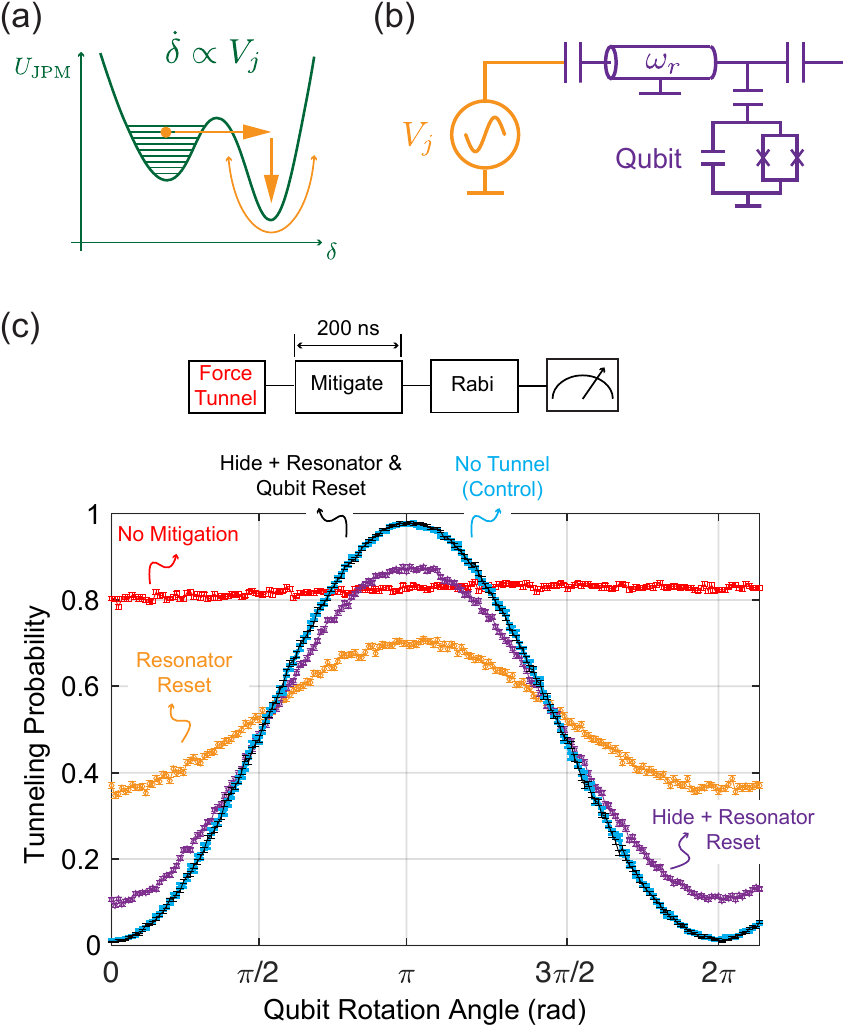}
     \caption{Characterizing and mitigating backaction induced by the JPM tunneling event. \textbf{(a)} Relaxation processes following a JPM tunneling event deposit an energy of order 100 photons on chip. \textbf{(b)} By the ac Josephson relation, the JPM is modeled as an effective voltage source $V_j$ that can excite both the resonator and qubit modes. \textbf{(c)} Rabi experiments preceded by a forced JPM tunneling event followed by various mitigation steps. The hide step is accomplished by biasing the qubit to a frequency where backaction from the forced tunneling event is minimal. With full mitigation (i.e. qubit hide bias plus resonator and qubit reset), we recover all but 0.2\% of the measurement fidelity compared to the experiment with no forced JPM tunneling event.  \label{fig:fig_7}}
\end{figure}

JPM tunneling events deposit an energy of order 100 photons on chip as the phase particle relaxes to the global minimum of the potential \cite{McDermott2005, Opremcak2018} [see Fig. \ref{fig:fig_7}{(a)}]. The associated transient contains spectral components at the frequencies of the readout resonator and the qubit, and as a result can transfer energy to these modes [Fig. \ref{fig:fig_7}{(b)}]. It is therefore critical to characterize the backaction and crosstalk associated with JPM tunneling events. 

We begin with a study of JPM-induced backaction using the Rabi experiment described in Fig. \ref{fig:fig_7}{(c)}. Prior to the qubit drive pulse, we force a tunneling event in the JPM and perform a deterministic reset of the JPM in the left well. In the absence of mitigation, the Rabi scan yields a nearly constant tunneling probability of 80\% as a function of the qubit rotation angle, indicating severe corruption of the qubit and the readout resonator by the JPM tunneling event. Next, we perform JPM-assisted resonator reset prior to the Rabi experiment as a potential mitigation strategy. Namely, we bias the JPM into resonance with the readout cavity for 100~ns as a means to deplete the cavity of photons released by the JPM tunneling event \cite{Opremcak2018}. With resonator reset, we recover Rabi oscillations with low visibility $\sim 30\%$. In a further refinement, we adjust the bias point of the qubit during the JPM tunneling event from 5.1~GHz down to 4.4~GHz in order to minimize the spectral content of the tunneling transient at the qubit frequency; we refer to this as a \textit{hide bias} step. By concatenating the hide bias step with resonator reset, we obtain a Rabi visibility $\sim75\%$. Finally, we append a JPM-assisted qubit reset step to the end of the mitigation sequence. With full mitigation, we recover all but 0.2\% of the measurement fidelity compared to the situation with no forced JPM tunneling event. The resonator and qubit reset steps take a combined time of 200~ns (see Appendix \ref{sec:resonator_qubit_reset}).

\begin{figure*}[t!]
    \includegraphics[scale=1.0]{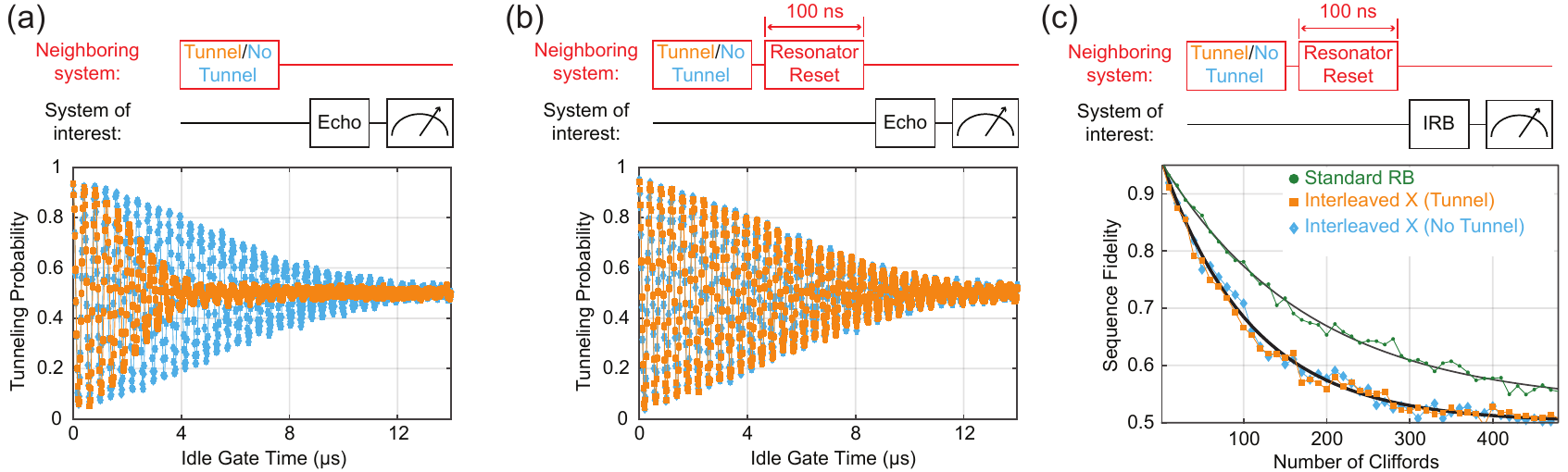}
     \caption{Characterizing and mitigating crosstalk induced by the JPM tunneling event. \textbf{(a)} Spin echo data taken on the q1-j1 system on chip \#1 with and without a prior forced tunneling event on the q2-j2 system. The spin-echo gate sequence is $X/2-\text{Idle}/2-Y-\text{Idle}/2-X/2$. We observe a factor of 2.6 reduction in the Gaussian decay envelope of the spin-echo fringes with respect to the control experiment with no forced tunneling event. \textbf{(b)} As in part (a), but with resonator reset on system q2-j2 following the forced tunneling event. We recover identical spin-echo fringes with respect to the control. \textbf{(c)} Interleaved randomized benchmarking (IRB) experiment to quantify the performance of our crosstalk mitigation strategy. We measure identical IRB gate fidelities for the tunnel and no tunnel cases following resonator reset on q2-j2 (see Table \ref{tab:tab3} for further detail).   \label{fig:fig_8}}
\end{figure*}

We characterize JPM-induced crosstalk to the unmeasured qubit by performing a spin-echo experiment on one qubit following a forced JPM tunneling event on the neighboring qubit-JPM pair [see Fig. \ref{fig:fig_8}{(a)}]. We use spin-echo to probe qubit coherence as opposed to a conventional Ramsey experiment in order to suppress the contribution to dephasing from low-frequency $1/f$ magnetic flux noise \cite{Martinis2003, Bylander2011, Hutchings2017}. We measure a factor of 2.6 reduction in the Gaussian decay time of the spin-echo fringes with respect to our control experiment \footnote{\textcolor{black}{The spin-echo decay envelope obtained from modeling photon shot noise dephasing using our circuit parameters is well described by a Gaussian function of the idle gate time}}, indicating the presence of unwanted crosstalk between systems. We speculate that the enhanced dephasing is due to spurious photon occupation in the measurement resonator of the tunneled JPM, leading to photon shot noise dephasing of the neighboring qubit via parasitic coupling \cite{Sears2012, Yan2016, Yan2018}. To test this hypothesis, we add a resonator reset step following the forced tunneling event as shown in Fig. \ref{fig:fig_8}{(b)}. With resonator reset, we recover identical spin-echo fringes with respect to the control experiment. To confirm that resonator reset fully mitigates crosstalk of the JPM-based measurement, we use IRB to quantify single-qubit gate error with and without a prior forced JPM tunneling event in the neighboring qubit-JPM system [Fig. \ref{fig:fig_8}{(c)}] \cite{Magesan2012}. With resonator reset following the JPM tunneling event, we measure identical interleaved gate fidelities for the tunnel and no tunnel cases, as summarized in Table \ref{tab:tab3}.

\begin{table}[h]
\begin{tabular}{ |P{2cm}|P{2.5cm}|P{2.5cm}| }
 \hline
 Interleaved Gate & Gate Fidelity (Tunnel) & Gate Fidelity (No Tunnel)\\
\hline
 $X$  & 99.8 $\pm$ 0.3\%  & 99.8 $\pm$ 0.2\%\\ 
$X/2$  & 99.9 $\pm$ 0.3\%  & 99.9 $\pm$ 0.2\%\\ 
$I$   & 99.9 $\pm$ 0.1\%  & 99.9 $\pm$ 0.1\%\\ 
 \hline
\end{tabular}
\caption{Interleaved randomized benchmarking results for the crosstalk experiments described in Section \ref{sec:level5}. \textcolor{black}{Each of the interleaved gates reported here has a total duration of 15 ns.} \label{tab:tab3} }
\end{table}

\begin{figure}[t!]
    \includegraphics[scale=1.0]{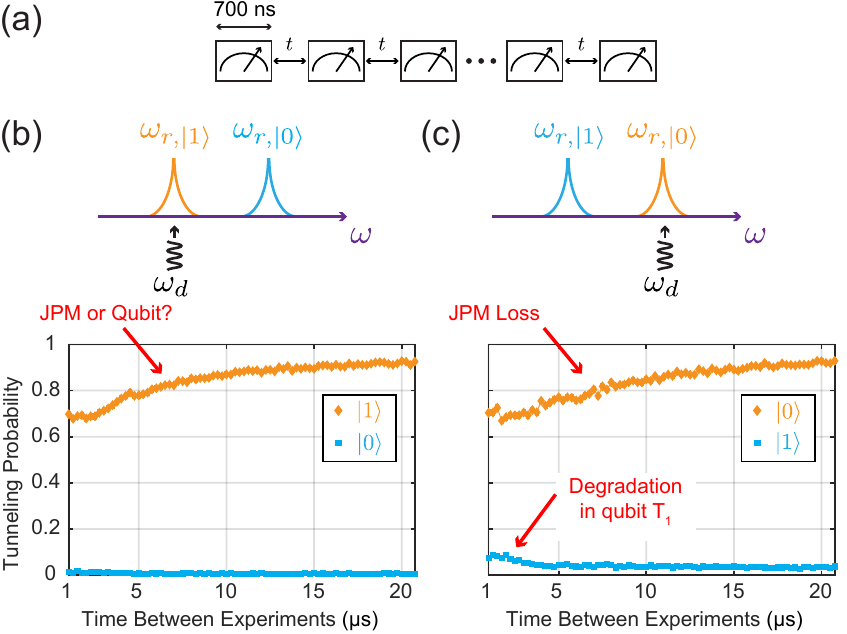}
     \caption{Dependence of JPM-based measurement fidelity on repetition rate. \textbf{(a)} Timing sequence of the experiment to probe sensitivity to measurement repetition rate. Each measurement takes 700~ns from start to finish, including resonator and qubit reset. \textbf{(b)} JPM-detection of the qubit $\ket{0}$ and $\ket{1}$ states versus interval between measurements. \textbf{(c)} As in (b), but with the qubit $\ket{0}$ state mapped to the bright pointer (i.e. $\omega_d=\omega_{r,\ket{0}}$). Parts (b) and (c) indicate a degradation in both JPM detection efficiency and qubit $T_1$ as the repetition rate is increased, with the former playing the dominant role.  \label{fig:fig_9}}
\end{figure}

To implement a practical error-corrected superconducting quantum computer based on the two-dimensional surface code, measurement repetition rates of order $1~\text{MHz}$ will be required \cite{Andersen2019, Andersen2020}. For this reason, we analyze the dependence of JPM-based measurement fidelity on the time between experiments using the measurement sequence depicted in Fig. \ref{fig:fig_9}{(a)}. We find that as the time between experiments decreases, the fidelity $P(1|1)$ with which we detect the bright pointer state degrades, with a characteristic time for recovery of fidelity of 13~$\mu$s [see Fig. \ref{fig:fig_9}{(b)}]. We speculate that the degradation in fidelity is due to enhanced loss in both the qubit and the JPM at high measurement repetition rates. To separately examine the contributions of the JPM and the qubit to the loss of fidelity, we switch the roles of the bright and dark pointer states as shown in Fig. \ref{fig:fig_9}{(c)}. With the qubit $\ket{1}$ state mapped to the dark cavity pointer, the measurement fidelity is insensitive to enhanced loss in the JPM, since an elevated JPM relaxation rate would preserve the correct outcome for measurement of the dark pointer state (namely, no tunneling event). However, in this case we do see enhanced $P(0|1)$ for measurement duty cycles below 5~$\mu$s, indicating a contribution to infidelity either from enhanced qubit relaxation or from qubit initialization errors. Similarly, when we map the qubit $\ket{0}$ state to the bright cavity pointer, the tunneling probability $P(0|0)$ is insensitive to qubit loss and dominated by enhanced loss in the JPM element that prevents mapping of the bright pointer state to a tunneling event.  We conclude that the enhanced measurement infidelity observed at high repetition rate is dominated by loss in the JPM, with a small contribution from increased qubit errors at the highest repetition rates $> 200$~kHz. While the physics that drives this degradation in fidelity is not presently understood, we speculate that the enhanced loss in both the qubit and the JPM is mediated by the transfer of energy released in the tunneling event to nonequilibrium quasiparticles \cite{Lenander2011, Wenner2013} or to dielectric two-level states (TLS) in the lossy bulk oxides of the JPM or in the surface oxides of the qubit. Possible mitigation strategies to preserve measurement fidelity at repetition rates approaching 1~MHz include incorporation of quasiparticle traps into the circuit \cite{Patel2017, Riwar2019} or a modification of the JPM energy landscape to reduce the energy released by the tunneling event.

\section{Conclusion\label{sec:level6}}

We have developed and characterized a fast, accurate state measurement technique for superconducting qubits using on-chip microwave photon counters. Our technique provides access to the binary result of projective quantum measurement at the millikelvin stage of a dilution refrigerator; furthermore, it eliminates the need for nonreciprocal circuit components between the qubit and the measurement apparatus \cite{Opremcak2018}. While our achieved raw single-shot measurement fidelity $>98\%$ already compares favorably with the current state of the art \cite{Arute2019}, straightforward improvements in pointer state preparation and suppression of qubit relaxation and initialization errors should push raw single-shot measurement fidelity beyond 99\%. 
Our study of achievable measurement repetition rate revealed an anomalous source of loss associated with JPM tunneling events; this topic merits further investigation. We anticipate that straightforward modifications to our circuit design will provide a path to higher measurement repetition rates.

The physical footprint of the JPM is well matched to the dimensions of the qubit, so that it would be straightforward to integrate a single JPM element with every qubit in a large-scale multiqubit processor; in such an architecture, each cell in the array would require one additional flux bias line for JPM control. Microwave-based readout of the classical flux state of the JPM is amenable to multiplexing for the efficient measurement of large multiqubit arrays with low hardware overhead; alternatively, it is possible to encode the flux state of the JPM in a propagating fluxon \cite{Herr2007, Fedorov2014, Howington2019} that could then be passed to a proximal classical Josephson digital circuit for error monitoring of the qubit array, postprocessing of the measurement results, and low-latency feedback. Combined with digital approaches to coherent control \cite{McDermott2014, Leonard2019}, this approach to measurement could form the basis for a scalable quantum--classical interface for next-generation superconducting qubit arrays \cite{McDermott2018}.

\begin{acknowledgments}
We thank Mostafa Khezri, Zijun Chen, Ofer Naaman, and John Martinis for stimulating discussions. We thank Tom McJunkin for assistance with device imaging. Portions of this work were supported by Google. R.M and B.L.T.P acknowledge funding from the National Science Foundation under Grants QIS-1720304 and QIS-1720312, respectively. A.O. was additionally funded by NSF award DMR-1747426. The authors gratefully acknowledge use of facilities and instrumentation at the UW-Madison Wisconsin Centers for Nanoscale Technology  (wcnt.wisc.edu), which is partially supported by the NSF through the University of Wisconsin Materials Research Science and Engineering Center (DMR-1720415).
\end{acknowledgments}
 
\appendix

\section{Sample Fabrication\label{sec:sample_fabrication}}
These samples were fabricated on a high-resistivity ($\gtrsim10~\text{k}\Omega\text{-cm}$) silicon substrate with 100 crystal orientation. Prior to base layer deposition, the substrate is dipped in dilute (2\%) hydrofluoric acid for one minute to remove native oxide from the surface. We then load the substrate into a dc magnetron sputter tool and deposit a 70 nm-thick film of Nb. The first patterning step defines all Nb features including the control wiring, measurement resonators, qubit capacitors, and spiral inductors. This pattern is then transferred into the Nb using an inductively coupled plasma etcher with Cl$_2$/BCl$_3$ chemistry. 
Next, we pattern the sample for liftoff and deposit the insulator used for crossover wiring and parallel-plate capacitors. The 180 nm-thick film of SiO$_2$ is deposited using an electron beam evaporator at an oxygen partial pressure $P_{\text{O}_2} = 10^{-5}$ Torr. In the final photolithography step, we pattern the sample for counterelectrode liftoff. We then deposit a 200 nm-thick Al counterelectrode using an electron beam evaporator after performing an \textit{in situ} ion mill clean to ensure good metallic contact to the base wiring layer. Finally, the JPM and qubit junctions are formed using a Dolan-bridge process \cite{Dolan1977} involving an MMA/PMMA resist stack patterned using a 100~keV electron-beam writer. The Al-AlO$_\text{x}$-Al junctions are shadow evaporated in an electron beam evaporator following an \textit{in situ} ion mill clean. This completes the device. Circuit parameters for the chip are listed in Table \ref{fig:tab4} with component labels indicated in Fig. \ref{fig:fig_2}(b).

\section{Measurement Setup\label{sec:experimental_setup}}
\begin{figure*}[ht]
    \includegraphics[]{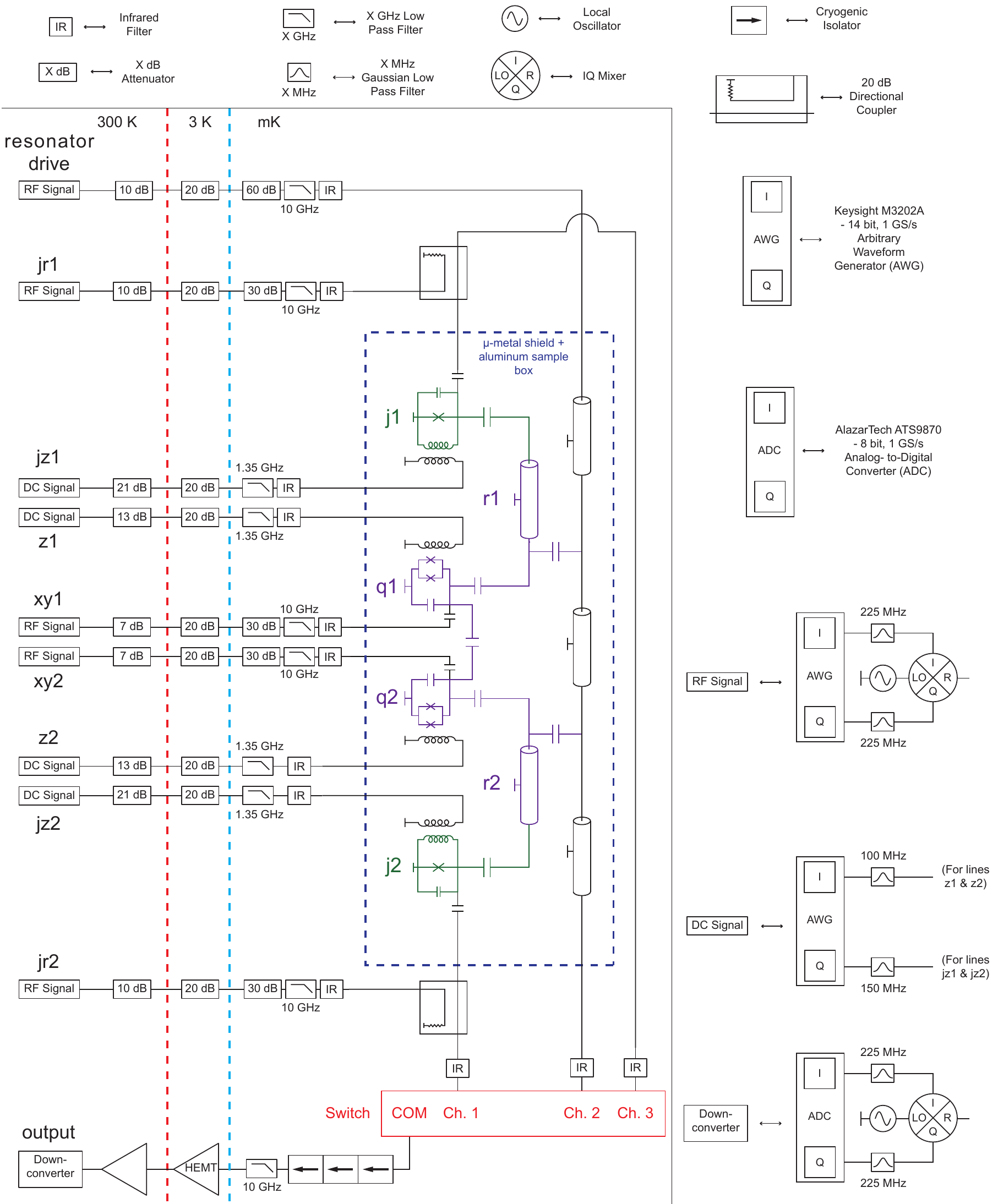}
     \caption{Wiring diagram of the experiment. Vertical dashed lines divide temperature stages. Circuit components are defined above and to the right. All components within the $\mu$-metal shield are made of non-magnetic materials. \label{fig:fig_13}}
\end{figure*}

The wiring diagram for our measurement setup is shown in Fig. \ref{fig:fig_13}. The waveforms for JPM readout (jr1/2), qubit excitation (xy1/2), and resonator drive are generated via single sideband mixing. Keysight M3202A arbitrary waveform generators (AWGs; 14 bit, 1 GS/s) produce intermediate frequency (IF) signals that are mixed with a local oscillator (LO) to generate shaped pulses at microwave frequencies. The qubit and JPM flux-bias waveforms (z1/2 and jz1/2, respectively) are directly synthesized using the AWGs. Signal rise times $\approx1-2~\text{ns}$ on the jz1/2 waveforms are critical to the success of the qubit measurement sequence [see Fig. \ref{fig:fig_4}{(a)}]. The state of the JPM is read out in reflection using a directional coupler. The reflected signal is passed through several stages of isolation and filtering prior to amplification by a high electron mobility transistor (HEMT) amplifier at the 3~K stage of the cryostat. Following additional room temperature amplification, the signal is sent to the RF port of an IQ mixer where it is down converted and digitized using an AlazarTech ATS9870 analog-to-digital converter (ADC; 8 bit, 1 GS/s). Further signal processing and thresholding are performed in software in order to extract the amplitude and phase of the reflected signal. The fidelity with which we measure the flux state of the JPM is better than 99.99\%; see Fig. \ref{fig:fig_3}{(b)}.


\section{Stark Calibration\label{sec:stark_calibration}}
\begin{figure}[t!]
    \includegraphics[]{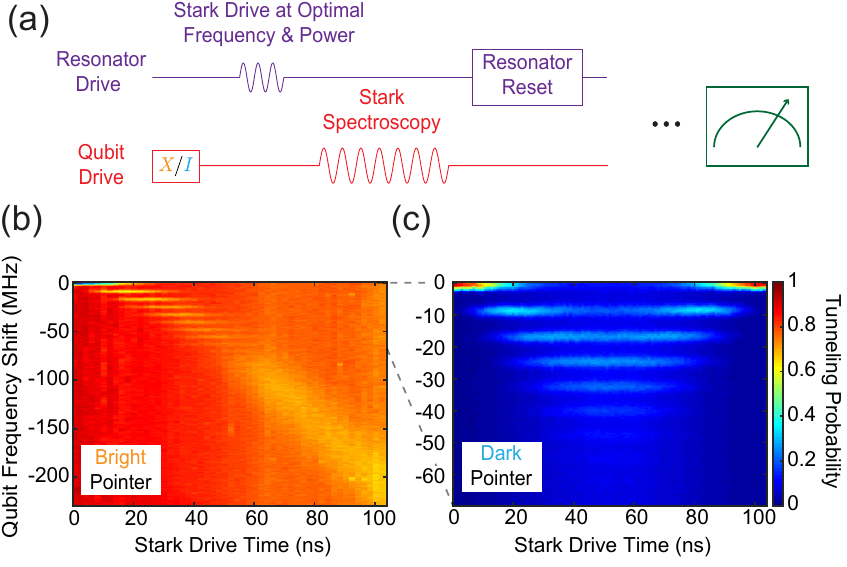}
     \caption{Stark calibration of pointer states. \textbf{(a)} Pulse sequence used for pointer-state calibration. These experiments were performed at the optimal resonator drive amplitude 0.885 arb. units (see Fig. \ref{fig:fig_5}). For details concerning resonator reset, see Appendix \ref{sec:resonator_qubit_reset}. \textbf{(b)} Qubit frequency shift versus Stark drive time for the bright pointer state. \textbf{(c)} As in (b), but for the dark pointer state.  \label{fig:fig_10}}
\end{figure}

We use the ac Stark effect \cite{Schuster2005, Schuster2007} to estimate photon occupation of the bright and dark pointer states; the pulse sequence is shown in Fig. \ref{fig:fig_10}{(a)}.
First, we prepare the qubit in $\ket{1}$ ($\ket{0}$) through the application of an $X$-gate ($I$-gate). Next, we drive the measurement resonator at the optimal frequency and power found in Fig. \ref{fig:fig_5} but for a variable amount of time, populating the measurement cavity with a mean number of photons $\bar{n}_r$. At the end of the Stark drive, a low-power, \textcolor{black}{500~ns-long} Stark spectroscopy pulse is applied to determine the qubit frequency shift $\Delta \omega_q \equiv \omega_q(\bar{n}_r) - \omega_q(\bar{n}_r=0)$. Because the photon lifetime in the readout cavity is relatively long $\sim1.5~\mu\text{s}$, $\bar{n}_r$ can be considered static on the timescale of the spectroscopy experiment. We then reset the resonator using the JPM to deplete the remaining photon occupation (see Appendix \ref{sec:resonator_qubit_reset}). Finally, we measure the qubit using the sequence described in Fig. \ref{fig:fig_4}. The results are shown in Fig. \ref{fig:fig_10}{(b, c)}. We find that the bright pointer state corresponds to a mean photon occupation of $\bar{n}_r \approx \Delta \omega_q/2\chi= 27~\text{photons}$, where $\Delta \omega_q/2\pi \approx -200~\text{MHz}$ at the optimal drive time ($t_d=105~\text{ns}$) and $2 \chi /2\pi  = 7.4~\text{MHz}$ is the Stark shift per photon. Similarly, the dark pointer acquires a maximum photon occupation $\bar{n}_r \approx 4~\text{photons}$ halfway through the drive pulse, but at the end of the resonator drive it returns to a state that is very close to vacuum. For this qubit operation point, the critical photon number $n_\text{crit} = (\Delta_{q, r}/g_{q, r})^2/4 \simeq 13~\text{photons}$. We note that these estimates of photon occupation neglect the effect of photon loss during the Stark spectroscopy pulse and the dependence of $\chi$ on $\bar{n}_r$.

\section{Fidelity Budget\label{sec:excess_one_state}}
\begin{figure}[t!]
    \includegraphics[]{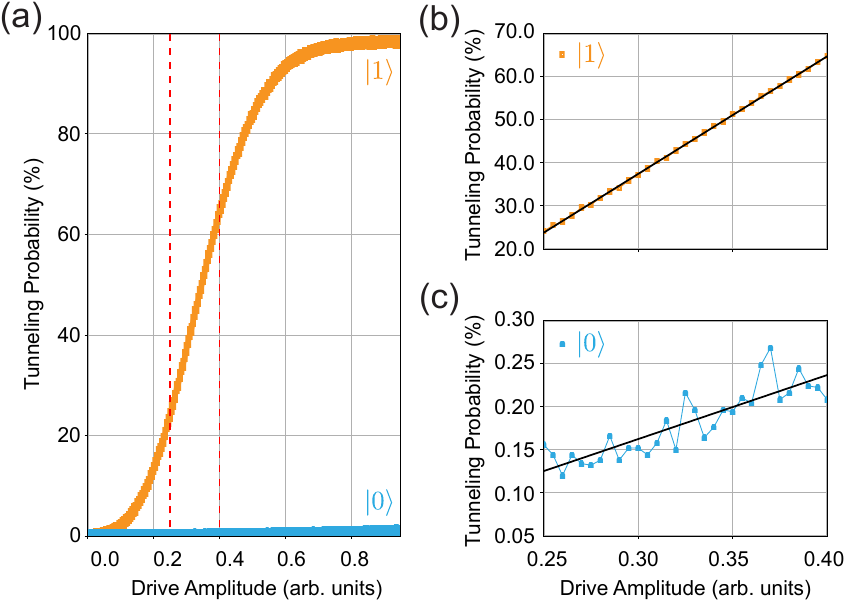}
     \caption{Estimating excess $\ket{1}$ population. \textbf{(a)} JPM tunneling probability versus resonator drive amplitude for qubits initialized in states $\ket{0}$ (blue) and $\ket{1}$ (orange). Based on our Stark calibration at the optimal drive amplitude 0.885 arb. units, we know that for drive amplitudes $\lesssim$ 0.4 arb. units, the maximum photon occupation of the dark pointer is less than one photon ($\ll n_\text{crit}$). Therefore, the dressed resonance corresponding to the qubit $\ket{0}$ state is well approximated by a linear mode during driven evolution. \textbf{(b)} Linear fits of JPM tunneling probability versus resonator drive amplitude over the range from 0.25-0.4 arb. units with the qubit prepared in $\ket{1}$. \textbf{(c)} As in (b), but with the qubit prepared in $\ket{0}$. \label{fig:fig_11}}
\end{figure}

The nonvanishing $P(1|0)$ contains contributions both from qubit initialization errors and from imperfect dark pointer state preparation. In order to separately quantify these errors, we performed a series of measurements following active reset of the qubit with resonator drive amplitude swept from its optimal value down to zero [Fig. \ref{fig:fig_11}{(a)}]; for comparison with Fig. \ref{fig:fig_10}, the calibration described in that figure was performed at a drive amplitude of 0.885 arb. units. As a result, we can be sure that for drive amplitude $\lesssim$ 0.4 arb. units, the maximum photon occupation of the dark pointer is less than one photon, which is much less than $n_\text{crit}$ over the entire course of driven evolution; at this level of cavity occupation, the dressed resonance corresponding to the qubit $\ket{0}$ state is well approximated by a linear mode. Therefore, we can attribute all of the tunneling at low resonator drive amplitude to excess $\ket{1}$ population alone, eliminating contributions caused by the Kerr nonlinearity of the resonator that occur at full drive strength. In Fig. \ref{fig:fig_11}{(b, c)}, we show linear fits to the data of Fig. \ref{fig:fig_11}{(a)} for resonator drive amplitudes ranging between 0.25-0.4 arb. units. The ratio of the slopes extracted from these fits gives an estimate of excess $\ket{1}$ population of 0.3\% for nominal preparation of the $\ket{0}$ state. We attribute the remaining contribution to $P(1|0)$ to imperfect dark pointer preparation, with infidelity 0.6\%.

\section{JPM-Assisted Resonator and Qubit Reset\label{sec:resonator_qubit_reset}}
\begin{figure}[t!]
    \includegraphics[]{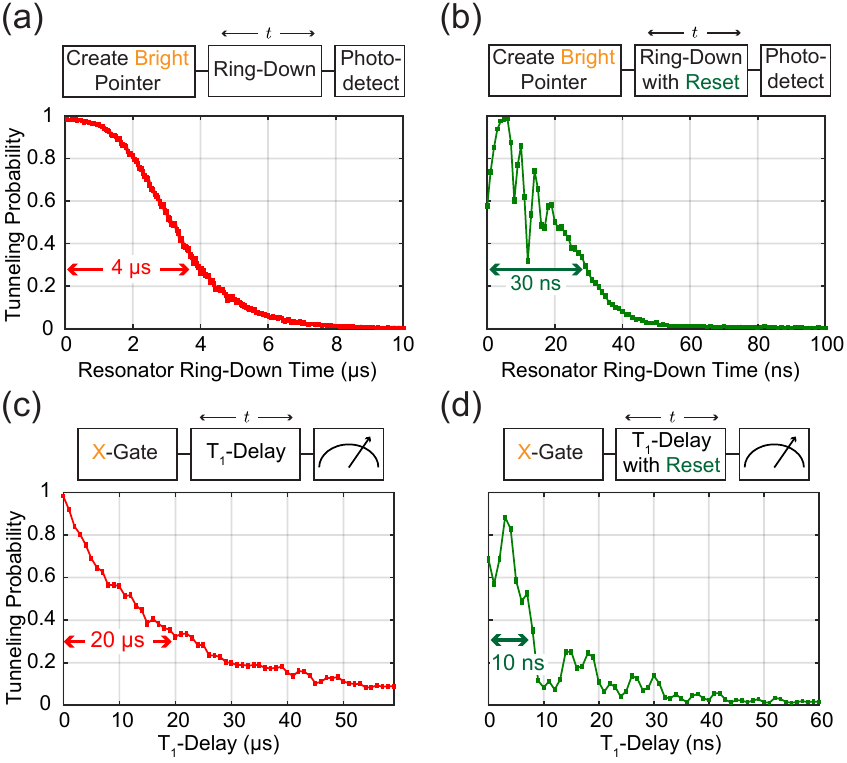}
     \caption{JPM-assisted resonator and qubit reset. \textbf{(a)} JPM photodetection of the bright pointer state after a variable ring-down delay. Passive resonator reset requires around 10~$\mu$s, which is too slow for the surface code cycle. \textbf{(b)} As in (a), but with the JPM biased into resonance with the resonator $\omega_r=\omega_j$ during the ring-down delay. Active resonator reset is performed in under 100~ns. \textbf{(c)} Qubit $T_1$ experiment. Passive qubit reset based on intrinsic relaxation processes requires a time of order 20~$\mu\text{s}$. \textbf{(d)} As in (c), but with the JPM and qubit biased into resonance with the measurement resonator ($\omega_r=\omega_j=\omega_q$) during the $T_1$-delay. Active qubit reset is performed in under $100~\text{ns}$. \label{fig:fig_12}}
\end{figure}

The intrinsic damping of the JPM provides an efficient method for the rapid reset of the resonator and qubit modes. This is accomplished by simply biasing the JPM into resonance with the mode of interest for a brief period of time. The data shown in Fig. \ref{fig:fig_12}{(a, b)} demonstrate reset of the measurement resonator. In Fig. \ref{fig:fig_12}(a) we plot JPM tunneling probability following photodetection of the bright pointer state after a variable ring-down delay. We observe that passive resonator reset requires $\simeq 10~\mu\text{s}$ to complete, a consequence of the high-$Q$ measurement resonator used in our design. To accelerate resonator reset, we bias the JPM into resonance with the measurement resonator during the ring-down delay, as shown in Fig. \ref{fig:fig_12}(b). With the JPM and resonator fully hybridized, the energy decay time of the mode is suppressed to around 10~ns, allowing for rapid on-demand depletion of the measurement resonator. We find that JPM-assisted resonator reset is accomplished in under 100~ns. 

We extend this idea to qubit reset in the experiments described in Fig. \ref{fig:fig_12}{(c, d)}. In each of these datasets, qubit $\ket{1}$ occupation is measured after the application of an $X$-gate followed by a variable delay. We find that passive reset based on qubit $T_1$ relaxation requires approximately 20~$\mu\text{s}$. However, when the JPM is biased into resonance with the qubit during reset, accurate qubit initialization is accomplished in under 100~ns. Throughout the experiments described in this manuscript, JPM-assisted qubit reset was used to suppress excess $\ket{1}$ state population from a baseline value of 4\% to 0.3\%.



\begin{center}
\begin{table*}[ht]
    \begin{tabular}{|P{1.5cm}|P{5cm}|P{2cm}|P{5cm}|}
    \hline
    Label & Description & Value & Method of Determination \\ \hline
     $g_{j_1}/2\pi$  & JPM-resonator coupling strength for j1-r1 & 62 MHz & JPM spectroscopy versus flux  \\ \hline
    $g_{j_2}/2\pi$ & JPM-resonator coupling strength for j2-r2 & 63 MHz & JPM spectroscopy versus flux  \\ \hline
    $g_{q_1}/2\pi$ & Qubit-resonator coupling strength for q1-r1 & 90 MHz & Qubit and resonator spectroscopy  \\ \hline
    $g_{q_2}/2\pi$ & Qubit-resonator coupling strength for q2-r2 & 92 MHz & Qubit and resonator spectroscopy  \\ \hline
    $\omega_{r_1}/2\pi$ & Bare frequency for resonator r1 & 5.693 GHz & High power resonator spectroscopy with j1 maximally detuned  \\ \hline
    $\omega_{r_2}/2\pi$ & Bare frequency for resonator r2 & 5.825 GHz & High power resonator spectroscopy with j2 maximally detuned  \\ \hline
    $\kappa_{r_1}$ & Total energy decay rate of resonator r1 & 1/(1.53 $\mu$s) & VNA measurements with j1 maximally detuned \\ \hline
    $\kappa_{r_2}$ & Total energy decay rate of resonator r2 & 1/(1.51 $\mu$s) & VNA measurements with j2 maximally detuned \\ \hline
    $g_{q_1,q_2}/2\pi$ & Qubit-qubit coupling strength & 16 MHz & Qubit spectroscopy about the avoided level crossing (degeneracy at 5.1 GHz) \\ \hline
    $T_{1,j}$ & Energy relaxation time of the JPM & 5 ns & VNA measurements with the JPM detuned from the resonator \\ \hline
    $L_{j}$ & Geometric inductance of the JPM & 1.3 nH & JPM spectroscopy versus flux \\ \hline
    $C_{j}$ & Self-capacitance of the JPM & 2.2 pF & JPM spectroscopy versus flux \\ \hline
	$C_{jr}$ & Reflection capacitor of the JPM & 33 fF & JPM spectroscopy versus flux \\ \hline
    $I_{0j}$ & Critical current of the JPM & 1.4 $\mu$A & JPM spectroscopy versus flux and 4-wire resistance measurements of cofabricated test junctions \\ \hline
    $M_{j}$ & Mutual inductance between the JPM and external bias circuitry & 4.8 pH & JPM spectroscopy versus flux \\ \hline
	$I_{0q}$ & Total critical current of the transmon dc SQUID loop & 43 nA & Qubit spectroscopy versus flux \\ \hline
    $M_{q}$ & Mutual inductance between the qubit and external bias circuitry & 1.4 pH & Resonator spectroscopy versus qubit flux \\ \hline
	$\eta/2\pi$ & Qubit anharmonicity & -225 MHz & Qubit spectroscopy of \textcolor{black}{the $\ket{0} \rightarrow \ket{1}$ and $\ket{1} \rightarrow \ket{2}$ transitions}  \\ \hline
    $C_{\text{xy}}$ & Qubit excitation capacitance & 40 aF & Sonnet simulation \\ \hline
    \end{tabular}
\caption{Circuit parameters for chip \#1. Labels can be found in Fig. \ref{fig:fig_2}{(a, b)}. \label{fig:tab4}}
\end{table*}
\end{center}

\clearpage
\bibliographystyle{apsrev4-1}
\bibliography{opremcak_refs}

\begin{thebibliography}{54}%
\makeatletter
\providecommand \@ifxundefined [1]{%
 \@ifx{#1\undefined}
}%
\providecommand \@ifnum [1]{%
 \ifnum #1\expandafter \@firstoftwo
 \else \expandafter \@secondoftwo
 \fi
}%
\providecommand \@ifx [1]{%
 \ifx #1\expandafter \@firstoftwo
 \else \expandafter \@secondoftwo
 \fi
}%
\providecommand \natexlab [1]{#1}%
\providecommand \enquote  [1]{``#1''}%
\providecommand \bibnamefont  [1]{#1}%
\providecommand \bibfnamefont [1]{#1}%
\providecommand \citenamefont [1]{#1}%
\providecommand \href@noop [0]{\@secondoftwo}%
\providecommand \href [0]{\begingroup \@sanitize@url \@href}%
\providecommand \@href[1]{\@@startlink{#1}\@@href}%
\providecommand \@@href[1]{\endgroup#1\@@endlink}%
\providecommand \@sanitize@url [0]{\catcode `\\12\catcode `\$12\catcode
  `\&12\catcode `\#12\catcode `\^12\catcode `\_12\catcode `\%12\relax}%
\providecommand \@@startlink[1]{}%
\providecommand \@@endlink[0]{}%
\providecommand \url  [0]{\begingroup\@sanitize@url \@url }%
\providecommand \@url [1]{\endgroup\@href {#1}{\urlprefix }}%
\providecommand \urlprefix  [0]{URL }%
\providecommand \Eprint [0]{\href }%
\providecommand \doibase [0]{http://dx.doi.org/}%
\providecommand \selectlanguage [0]{\@gobble}%
\providecommand \bibinfo  [0]{\@secondoftwo}%
\providecommand \bibfield  [0]{\@secondoftwo}%
\providecommand \translation [1]{[#1]}%
\providecommand \BibitemOpen [0]{}%
\providecommand \bibitemStop [0]{}%
\providecommand \bibitemNoStop [0]{.\EOS\space}%
\providecommand \EOS [0]{\spacefactor3000\relax}%
\providecommand \BibitemShut  [1]{\csname bibitem#1\endcsname}%
\let\auto@bib@innerbib\@empty
\bibitem [{\citenamefont {Fowler}\ \emph {et~al.}(2012)\citenamefont {Fowler},
  \citenamefont {Mariantoni}, \citenamefont {Martinis},\ and\ \citenamefont
  {Cleland}}]{Fowler2012}%
  \BibitemOpen
  \bibfield  {author} {\bibinfo {author} {\bibfnamefont {A.~G.}\ \bibnamefont
  {Fowler}}, \bibinfo {author} {\bibfnamefont {M.}~\bibnamefont {Mariantoni}},
  \bibinfo {author} {\bibfnamefont {J.~M.}\ \bibnamefont {Martinis}}, \ and\
  \bibinfo {author} {\bibfnamefont {A.~N.}\ \bibnamefont {Cleland}},\ }\href
  {\doibase 10.1103/PhysRevA.86.032324} {\bibfield  {journal} {\bibinfo
  {journal} {Phys. Rev. A}\ }\textbf {\bibinfo {volume} {86}},\ \bibinfo
  {pages} {032324} (\bibinfo {year} {2012})}\BibitemShut {NoStop}%
\bibitem [{\citenamefont {McDermott}\ \emph {et~al.}(2018)\citenamefont
  {McDermott}, \citenamefont {Vavilov}, \citenamefont {Plourde}, \citenamefont
  {Wilhelm}, \citenamefont {Liebermann}, \citenamefont {Mukhanov},\ and\
  \citenamefont {Ohki}}]{McDermott2018}%
  \BibitemOpen
  \bibfield  {author} {\bibinfo {author} {\bibfnamefont {R.}~\bibnamefont
  {McDermott}}, \bibinfo {author} {\bibfnamefont {M.~G.}\ \bibnamefont
  {Vavilov}}, \bibinfo {author} {\bibfnamefont {B.~L.~T.}\ \bibnamefont
  {Plourde}}, \bibinfo {author} {\bibfnamefont {F.~K.}\ \bibnamefont
  {Wilhelm}}, \bibinfo {author} {\bibfnamefont {P.~J.}\ \bibnamefont
  {Liebermann}}, \bibinfo {author} {\bibfnamefont {O.~A.}\ \bibnamefont
  {Mukhanov}}, \ and\ \bibinfo {author} {\bibfnamefont {T.~A.}\ \bibnamefont
  {Ohki}},\ }\href {\doibase 10.1088/2058-9565/aaa3a0} {\bibfield  {journal}
  {\bibinfo  {journal} {Quantum Sci. Technol.}\ }\textbf {\bibinfo {volume}
  {3}},\ \bibinfo {pages} {024004} (\bibinfo {year} {2018})}\BibitemShut
  {NoStop}%
\bibitem [{\citenamefont {Jeffrey}\ \emph {et~al.}(2014)\citenamefont
  {Jeffrey}, \citenamefont {Sank}, \citenamefont {Mutus}, \citenamefont
  {White}, \citenamefont {Kelly}, \citenamefont {Barends}, \citenamefont
  {Chen}, \citenamefont {Chen}, \citenamefont {Chiaro}, \citenamefont
  {Dunsworth}, \citenamefont {Megrant}, \citenamefont {O'Malley}, \citenamefont
  {Neill}, \citenamefont {Roushan}, \citenamefont {Vainsencher}, \citenamefont
  {Wenner}, \citenamefont {Cleland},\ and\ \citenamefont
  {Martinis}}]{Jeffrey2014}%
  \BibitemOpen
  \bibfield  {author} {\bibinfo {author} {\bibfnamefont {E.}~\bibnamefont
  {Jeffrey}}, \bibinfo {author} {\bibfnamefont {D.}~\bibnamefont {Sank}},
  \bibinfo {author} {\bibfnamefont {J.~Y.}\ \bibnamefont {Mutus}}, \bibinfo
  {author} {\bibfnamefont {T.~C.}\ \bibnamefont {White}}, \bibinfo {author}
  {\bibfnamefont {J.}~\bibnamefont {Kelly}}, \bibinfo {author} {\bibfnamefont
  {R.}~\bibnamefont {Barends}}, \bibinfo {author} {\bibfnamefont
  {Y.}~\bibnamefont {Chen}}, \bibinfo {author} {\bibfnamefont {Z.}~\bibnamefont
  {Chen}}, \bibinfo {author} {\bibfnamefont {B.}~\bibnamefont {Chiaro}},
  \bibinfo {author} {\bibfnamefont {A.}~\bibnamefont {Dunsworth}}, \bibinfo
  {author} {\bibfnamefont {A.}~\bibnamefont {Megrant}}, \bibinfo {author}
  {\bibfnamefont {P.~J.~J.}\ \bibnamefont {O'Malley}}, \bibinfo {author}
  {\bibfnamefont {C.}~\bibnamefont {Neill}}, \bibinfo {author} {\bibfnamefont
  {P.}~\bibnamefont {Roushan}}, \bibinfo {author} {\bibfnamefont
  {A.}~\bibnamefont {Vainsencher}}, \bibinfo {author} {\bibfnamefont
  {J.}~\bibnamefont {Wenner}}, \bibinfo {author} {\bibfnamefont {A.~N.}\
  \bibnamefont {Cleland}}, \ and\ \bibinfo {author} {\bibfnamefont {J.~M.}\
  \bibnamefont {Martinis}},\ }\href {\doibase 10.1103/PhysRevLett.112.190504}
  {\bibfield  {journal} {\bibinfo  {journal} {Phys. Rev. Lett.}\ }\textbf
  {\bibinfo {volume} {112}},\ \bibinfo {pages} {190504} (\bibinfo {year}
  {2014})}\BibitemShut {NoStop}%
\bibitem [{\citenamefont {Walter}\ \emph {et~al.}(2017)\citenamefont {Walter},
  \citenamefont {Kurpiers}, \citenamefont {Gasparinetti}, \citenamefont
  {Magnard}, \citenamefont {Poto\ifmmode~\check{c}\else \v{c}\fi{}nik},
  \citenamefont {Salath\'e}, \citenamefont {Pechal}, \citenamefont {Mondal},
  \citenamefont {Oppliger}, \citenamefont {Eichler},\ and\ \citenamefont
  {Wallraff}}]{Walter2017}%
  \BibitemOpen
  \bibfield  {author} {\bibinfo {author} {\bibfnamefont {T.}~\bibnamefont
  {Walter}}, \bibinfo {author} {\bibfnamefont {P.}~\bibnamefont {Kurpiers}},
  \bibinfo {author} {\bibfnamefont {S.}~\bibnamefont {Gasparinetti}}, \bibinfo
  {author} {\bibfnamefont {P.}~\bibnamefont {Magnard}}, \bibinfo {author}
  {\bibfnamefont {A.}~\bibnamefont {Poto\ifmmode~\check{c}\else
  \v{c}\fi{}nik}}, \bibinfo {author} {\bibfnamefont {Y.}~\bibnamefont
  {Salath\'e}}, \bibinfo {author} {\bibfnamefont {M.}~\bibnamefont {Pechal}},
  \bibinfo {author} {\bibfnamefont {M.}~\bibnamefont {Mondal}}, \bibinfo
  {author} {\bibfnamefont {M.}~\bibnamefont {Oppliger}}, \bibinfo {author}
  {\bibfnamefont {C.}~\bibnamefont {Eichler}}, \ and\ \bibinfo {author}
  {\bibfnamefont {A.}~\bibnamefont {Wallraff}},\ }\href {\doibase
  10.1103/PhysRevApplied.7.054020} {\bibfield  {journal} {\bibinfo  {journal}
  {Phys. Rev. Applied}\ }\textbf {\bibinfo {volume} {7}},\ \bibinfo {pages}
  {054020} (\bibinfo {year} {2017})}\BibitemShut {NoStop}%
\bibitem [{\citenamefont {Heinsoo}\ \emph {et~al.}(2018)\citenamefont
  {Heinsoo}, \citenamefont {Andersen}, \citenamefont {Remm}, \citenamefont
  {Krinner}, \citenamefont {Walter}, \citenamefont {Salath\'e}, \citenamefont
  {Gasparinetti}, \citenamefont {Besse}, \citenamefont
  {Poto\ifmmode~\check{c}\else \v{c}\fi{}nik}, \citenamefont {Wallraff},\ and\
  \citenamefont {Eichler}}]{Heinsoo2018}%
  \BibitemOpen
  \bibfield  {author} {\bibinfo {author} {\bibfnamefont {J.}~\bibnamefont
  {Heinsoo}}, \bibinfo {author} {\bibfnamefont {C.~K.}\ \bibnamefont
  {Andersen}}, \bibinfo {author} {\bibfnamefont {A.}~\bibnamefont {Remm}},
  \bibinfo {author} {\bibfnamefont {S.}~\bibnamefont {Krinner}}, \bibinfo
  {author} {\bibfnamefont {T.}~\bibnamefont {Walter}}, \bibinfo {author}
  {\bibfnamefont {Y.}~\bibnamefont {Salath\'e}}, \bibinfo {author}
  {\bibfnamefont {S.}~\bibnamefont {Gasparinetti}}, \bibinfo {author}
  {\bibfnamefont {J.-C.}\ \bibnamefont {Besse}}, \bibinfo {author}
  {\bibfnamefont {A.}~\bibnamefont {Poto\ifmmode~\check{c}\else
  \v{c}\fi{}nik}}, \bibinfo {author} {\bibfnamefont {A.}~\bibnamefont
  {Wallraff}}, \ and\ \bibinfo {author} {\bibfnamefont {C.}~\bibnamefont
  {Eichler}},\ }\href {\doibase 10.1103/PhysRevApplied.10.034040} {\bibfield
  {journal} {\bibinfo  {journal} {Phys. Rev. Applied}\ }\textbf {\bibinfo
  {volume} {10}},\ \bibinfo {pages} {034040} (\bibinfo {year}
  {2018})}\BibitemShut {NoStop}%
\bibitem [{\citenamefont {Mutus}\ \emph {et~al.}(2014)\citenamefont {Mutus},
  \citenamefont {White}, \citenamefont {Barends}, \citenamefont {Chen},
  \citenamefont {Chen}, \citenamefont {Chiaro}, \citenamefont {Dunsworth},
  \citenamefont {Jeffrey}, \citenamefont {Kelly}, \citenamefont {Megrant},
  \citenamefont {Neill}, \citenamefont {O'Malley}, \citenamefont {Roushan},
  \citenamefont {Sank}, \citenamefont {Vainsencher}, \citenamefont {Wenner},
  \citenamefont {Sundqvist}, \citenamefont {Cleland},\ and\ \citenamefont
  {Martinis}}]{Mutus2014}%
  \BibitemOpen
  \bibfield  {author} {\bibinfo {author} {\bibfnamefont {J.~Y.}\ \bibnamefont
  {Mutus}}, \bibinfo {author} {\bibfnamefont {T.~C.}\ \bibnamefont {White}},
  \bibinfo {author} {\bibfnamefont {R.}~\bibnamefont {Barends}}, \bibinfo
  {author} {\bibfnamefont {Y.}~\bibnamefont {Chen}}, \bibinfo {author}
  {\bibfnamefont {Z.}~\bibnamefont {Chen}}, \bibinfo {author} {\bibfnamefont
  {B.}~\bibnamefont {Chiaro}}, \bibinfo {author} {\bibfnamefont
  {A.}~\bibnamefont {Dunsworth}}, \bibinfo {author} {\bibfnamefont
  {E.}~\bibnamefont {Jeffrey}}, \bibinfo {author} {\bibfnamefont
  {J.}~\bibnamefont {Kelly}}, \bibinfo {author} {\bibfnamefont
  {A.}~\bibnamefont {Megrant}}, \bibinfo {author} {\bibfnamefont
  {C.}~\bibnamefont {Neill}}, \bibinfo {author} {\bibfnamefont {P.~J.~J.}\
  \bibnamefont {O'Malley}}, \bibinfo {author} {\bibfnamefont {P.}~\bibnamefont
  {Roushan}}, \bibinfo {author} {\bibfnamefont {D.}~\bibnamefont {Sank}},
  \bibinfo {author} {\bibfnamefont {A.}~\bibnamefont {Vainsencher}}, \bibinfo
  {author} {\bibfnamefont {J.}~\bibnamefont {Wenner}}, \bibinfo {author}
  {\bibfnamefont {K.~M.}\ \bibnamefont {Sundqvist}}, \bibinfo {author}
  {\bibfnamefont {A.~N.}\ \bibnamefont {Cleland}}, \ and\ \bibinfo {author}
  {\bibfnamefont {J.~M.}\ \bibnamefont {Martinis}},\ }\href {\doibase
  10.1063/1.4886408} {\bibfield  {journal} {\bibinfo  {journal} {Appl. Phys.
  Lett.}\ }\textbf {\bibinfo {volume} {104}},\ \bibinfo {pages} {263513}
  (\bibinfo {year} {2014})}\BibitemShut {NoStop}%
\bibitem [{\citenamefont {Macklin}\ \emph {et~al.}(2015)\citenamefont
  {Macklin}, \citenamefont {O{\textquoteright}Brien}, \citenamefont {Hover},
  \citenamefont {Schwartz}, \citenamefont {Bolkhovsky}, \citenamefont {Zhang},
  \citenamefont {Oliver},\ and\ \citenamefont {Siddiqi}}]{Macklin2015}%
  \BibitemOpen
  \bibfield  {author} {\bibinfo {author} {\bibfnamefont {C.}~\bibnamefont
  {Macklin}}, \bibinfo {author} {\bibfnamefont {K.}~\bibnamefont
  {O{\textquoteright}Brien}}, \bibinfo {author} {\bibfnamefont
  {D.}~\bibnamefont {Hover}}, \bibinfo {author} {\bibfnamefont {M.~E.}\
  \bibnamefont {Schwartz}}, \bibinfo {author} {\bibfnamefont {V.}~\bibnamefont
  {Bolkhovsky}}, \bibinfo {author} {\bibfnamefont {X.}~\bibnamefont {Zhang}},
  \bibinfo {author} {\bibfnamefont {W.~D.}\ \bibnamefont {Oliver}}, \ and\
  \bibinfo {author} {\bibfnamefont {I.}~\bibnamefont {Siddiqi}},\ }\href
  {\doibase 10.1126/science.aaa8525} {\bibfield  {journal} {\bibinfo  {journal}
  {Science}\ }\textbf {\bibinfo {volume} {350}},\ \bibinfo {pages} {307}
  (\bibinfo {year} {2015})}\BibitemShut {NoStop}%
\bibitem [{\citenamefont {Sliwa}\ \emph {et~al.}(2015)\citenamefont {Sliwa},
  \citenamefont {Hatridge}, \citenamefont {Narla}, \citenamefont {Shankar},
  \citenamefont {Frunzio}, \citenamefont {Schoelkopf},\ and\ \citenamefont
  {Devoret}}]{Sliwa2015}%
  \BibitemOpen
  \bibfield  {author} {\bibinfo {author} {\bibfnamefont {K.~M.}\ \bibnamefont
  {Sliwa}}, \bibinfo {author} {\bibfnamefont {M.}~\bibnamefont {Hatridge}},
  \bibinfo {author} {\bibfnamefont {A.}~\bibnamefont {Narla}}, \bibinfo
  {author} {\bibfnamefont {S.}~\bibnamefont {Shankar}}, \bibinfo {author}
  {\bibfnamefont {L.}~\bibnamefont {Frunzio}}, \bibinfo {author} {\bibfnamefont
  {R.~J.}\ \bibnamefont {Schoelkopf}}, \ and\ \bibinfo {author} {\bibfnamefont
  {M.~H.}\ \bibnamefont {Devoret}},\ }\href {\doibase
  10.1103/PhysRevX.5.041020} {\bibfield  {journal} {\bibinfo  {journal} {Phys.
  Rev. X}\ }\textbf {\bibinfo {volume} {5}},\ \bibinfo {pages} {041020}
  (\bibinfo {year} {2015})}\BibitemShut {NoStop}%
\bibitem [{\citenamefont {Lecocq}\ \emph {et~al.}(2017)\citenamefont {Lecocq},
  \citenamefont {Ranzani}, \citenamefont {Peterson}, \citenamefont {Cicak},
  \citenamefont {Simmonds}, \citenamefont {Teufel},\ and\ \citenamefont
  {Aumentado}}]{Lecocq2017}%
  \BibitemOpen
  \bibfield  {author} {\bibinfo {author} {\bibfnamefont {F.}~\bibnamefont
  {Lecocq}}, \bibinfo {author} {\bibfnamefont {L.}~\bibnamefont {Ranzani}},
  \bibinfo {author} {\bibfnamefont {G.~A.}\ \bibnamefont {Peterson}}, \bibinfo
  {author} {\bibfnamefont {K.}~\bibnamefont {Cicak}}, \bibinfo {author}
  {\bibfnamefont {R.~W.}\ \bibnamefont {Simmonds}}, \bibinfo {author}
  {\bibfnamefont {J.~D.}\ \bibnamefont {Teufel}}, \ and\ \bibinfo {author}
  {\bibfnamefont {J.}~\bibnamefont {Aumentado}},\ }\href {\doibase
  10.1103/PhysRevApplied.7.024028} {\bibfield  {journal} {\bibinfo  {journal}
  {Phys. Rev. Applied}\ }\textbf {\bibinfo {volume} {7}},\ \bibinfo {pages}
  {024028} (\bibinfo {year} {2017})}\BibitemShut {NoStop}%
\bibitem [{\citenamefont {Chapman}\ \emph {et~al.}(2017)\citenamefont
  {Chapman}, \citenamefont {Rosenthal}, \citenamefont {Kerckhoff},
  \citenamefont {Moores}, \citenamefont {Vale}, \citenamefont {Mates},
  \citenamefont {Hilton}, \citenamefont {Lalumi\`ere}, \citenamefont {Blais},\
  and\ \citenamefont {Lehnert}}]{Chapman2017}%
  \BibitemOpen
  \bibfield  {author} {\bibinfo {author} {\bibfnamefont {B.~J.}\ \bibnamefont
  {Chapman}}, \bibinfo {author} {\bibfnamefont {E.~I.}\ \bibnamefont
  {Rosenthal}}, \bibinfo {author} {\bibfnamefont {J.}~\bibnamefont
  {Kerckhoff}}, \bibinfo {author} {\bibfnamefont {B.~A.}\ \bibnamefont
  {Moores}}, \bibinfo {author} {\bibfnamefont {L.~R.}\ \bibnamefont {Vale}},
  \bibinfo {author} {\bibfnamefont {J.~A.~B.}\ \bibnamefont {Mates}}, \bibinfo
  {author} {\bibfnamefont {G.~C.}\ \bibnamefont {Hilton}}, \bibinfo {author}
  {\bibfnamefont {K.}~\bibnamefont {Lalumi\`ere}}, \bibinfo {author}
  {\bibfnamefont {A.}~\bibnamefont {Blais}}, \ and\ \bibinfo {author}
  {\bibfnamefont {K.~W.}\ \bibnamefont {Lehnert}},\ }\href {\doibase
  10.1103/PhysRevX.7.041043} {\bibfield  {journal} {\bibinfo  {journal} {Phys.
  Rev. X}\ }\textbf {\bibinfo {volume} {7}},\ \bibinfo {pages} {041043}
  (\bibinfo {year} {2017})}\BibitemShut {NoStop}%
\bibitem [{\citenamefont {Thorbeck}\ \emph {et~al.}(2017)\citenamefont
  {Thorbeck}, \citenamefont {Zhu}, \citenamefont {Leonard}, \citenamefont
  {Barends}, \citenamefont {Kelly}, \citenamefont {Martinis},\ and\
  \citenamefont {McDermott}}]{Thorbeck2017}%
  \BibitemOpen
  \bibfield  {author} {\bibinfo {author} {\bibfnamefont {T.}~\bibnamefont
  {Thorbeck}}, \bibinfo {author} {\bibfnamefont {S.}~\bibnamefont {Zhu}},
  \bibinfo {author} {\bibfnamefont {E.}~\bibnamefont {Leonard}}, \bibinfo
  {author} {\bibfnamefont {R.}~\bibnamefont {Barends}}, \bibinfo {author}
  {\bibfnamefont {J.}~\bibnamefont {Kelly}}, \bibinfo {author} {\bibfnamefont
  {J.~M.}\ \bibnamefont {Martinis}}, \ and\ \bibinfo {author} {\bibfnamefont
  {R.}~\bibnamefont {McDermott}},\ }\href {\doibase
  10.1103/PhysRevApplied.8.054007} {\bibfield  {journal} {\bibinfo  {journal}
  {Phys. Rev. Applied}\ }\textbf {\bibinfo {volume} {8}},\ \bibinfo {pages}
  {054007} (\bibinfo {year} {2017})}\BibitemShut {NoStop}%
\bibitem [{\citenamefont {Abdo}\ \emph {et~al.}(2019)\citenamefont {Abdo},
  \citenamefont {Bronn}, \citenamefont {Jinka}, \citenamefont {Olivadese},
  \citenamefont {C{\'o}rcoles}, \citenamefont {Adiga}, \citenamefont {Brink},
  \citenamefont {Lake}, \citenamefont {Wu}, \citenamefont {Pappas},\ and\
  \citenamefont {Chow}}]{Abdo2019}%
  \BibitemOpen
  \bibfield  {author} {\bibinfo {author} {\bibfnamefont {B.}~\bibnamefont
  {Abdo}}, \bibinfo {author} {\bibfnamefont {N.~T.}\ \bibnamefont {Bronn}},
  \bibinfo {author} {\bibfnamefont {O.}~\bibnamefont {Jinka}}, \bibinfo
  {author} {\bibfnamefont {S.}~\bibnamefont {Olivadese}}, \bibinfo {author}
  {\bibfnamefont {A.~D.}\ \bibnamefont {C{\'o}rcoles}}, \bibinfo {author}
  {\bibfnamefont {V.~P.}\ \bibnamefont {Adiga}}, \bibinfo {author}
  {\bibfnamefont {M.}~\bibnamefont {Brink}}, \bibinfo {author} {\bibfnamefont
  {R.~E.}\ \bibnamefont {Lake}}, \bibinfo {author} {\bibfnamefont
  {X.}~\bibnamefont {Wu}}, \bibinfo {author} {\bibfnamefont {D.~P.}\
  \bibnamefont {Pappas}}, \ and\ \bibinfo {author} {\bibfnamefont {J.~M.}\
  \bibnamefont {Chow}},\ }\href {\doibase 10.1038/s41467-019-11101-3}
  {\bibfield  {journal} {\bibinfo  {journal} {Nat. Commun.}\ }\textbf {\bibinfo
  {volume} {10}},\ \bibinfo {pages} {3154} (\bibinfo {year}
  {2019})}\BibitemShut {NoStop}%
\bibitem [{\citenamefont {Eddins}\ \emph {et~al.}(2019)\citenamefont {Eddins},
  \citenamefont {Kreikebaum}, \citenamefont {Toyli}, \citenamefont
  {Levenson-Falk}, \citenamefont {Dove}, \citenamefont {Livingston},
  \citenamefont {Levitan}, \citenamefont {Govia}, \citenamefont {Clerk},\ and\
  \citenamefont {Siddiqi}}]{Eddins2019}%
  \BibitemOpen
  \bibfield  {author} {\bibinfo {author} {\bibfnamefont {A.}~\bibnamefont
  {Eddins}}, \bibinfo {author} {\bibfnamefont {J.~M.}\ \bibnamefont
  {Kreikebaum}}, \bibinfo {author} {\bibfnamefont {D.~M.}\ \bibnamefont
  {Toyli}}, \bibinfo {author} {\bibfnamefont {E.~M.}\ \bibnamefont
  {Levenson-Falk}}, \bibinfo {author} {\bibfnamefont {A.}~\bibnamefont {Dove}},
  \bibinfo {author} {\bibfnamefont {W.~P.}\ \bibnamefont {Livingston}},
  \bibinfo {author} {\bibfnamefont {B.~A.}\ \bibnamefont {Levitan}}, \bibinfo
  {author} {\bibfnamefont {L.~C.~G.}\ \bibnamefont {Govia}}, \bibinfo {author}
  {\bibfnamefont {A.~A.}\ \bibnamefont {Clerk}}, \ and\ \bibinfo {author}
  {\bibfnamefont {I.}~\bibnamefont {Siddiqi}},\ }\href {\doibase
  10.1103/PhysRevX.9.011004} {\bibfield  {journal} {\bibinfo  {journal} {Phys.
  Rev. X}\ }\textbf {\bibinfo {volume} {9}},\ \bibinfo {pages} {011004}
  (\bibinfo {year} {2019})}\BibitemShut {NoStop}%
\bibitem [{\citenamefont {Blais}\ \emph {et~al.}(2004)\citenamefont {Blais},
  \citenamefont {Huang}, \citenamefont {Wallraff}, \citenamefont {Girvin},\
  and\ \citenamefont {Schoelkopf}}]{Blais2004}%
  \BibitemOpen
  \bibfield  {author} {\bibinfo {author} {\bibfnamefont {A.}~\bibnamefont
  {Blais}}, \bibinfo {author} {\bibfnamefont {R.-S.}\ \bibnamefont {Huang}},
  \bibinfo {author} {\bibfnamefont {A.}~\bibnamefont {Wallraff}}, \bibinfo
  {author} {\bibfnamefont {S.~M.}\ \bibnamefont {Girvin}}, \ and\ \bibinfo
  {author} {\bibfnamefont {R.~J.}\ \bibnamefont {Schoelkopf}},\ }\href
  {\doibase 10.1103/PhysRevA.69.062320} {\bibfield  {journal} {\bibinfo
  {journal} {Phys. Rev. A}\ }\textbf {\bibinfo {volume} {69}},\ \bibinfo
  {pages} {062320} (\bibinfo {year} {2004})}\BibitemShut {NoStop}%
\bibitem [{\citenamefont {Govia}\ \emph {et~al.}(2014)\citenamefont {Govia},
  \citenamefont {Pritchett}, \citenamefont {Xu}, \citenamefont {Plourde},
  \citenamefont {Vavilov}, \citenamefont {Wilhelm},\ and\ \citenamefont
  {McDermott}}]{Govia2014}%
  \BibitemOpen
  \bibfield  {author} {\bibinfo {author} {\bibfnamefont {L.~C.~G.}\
  \bibnamefont {Govia}}, \bibinfo {author} {\bibfnamefont {E.~J.}\ \bibnamefont
  {Pritchett}}, \bibinfo {author} {\bibfnamefont {C.}~\bibnamefont {Xu}},
  \bibinfo {author} {\bibfnamefont {B.~L.~T.}\ \bibnamefont {Plourde}},
  \bibinfo {author} {\bibfnamefont {M.~G.}\ \bibnamefont {Vavilov}}, \bibinfo
  {author} {\bibfnamefont {F.~K.}\ \bibnamefont {Wilhelm}}, \ and\ \bibinfo
  {author} {\bibfnamefont {R.}~\bibnamefont {McDermott}},\ }\href {\doibase
  10.1103/PhysRevA.90.062307} {\bibfield  {journal} {\bibinfo  {journal} {Phys.
  Rev. A}\ }\textbf {\bibinfo {volume} {90}},\ \bibinfo {pages} {062307}
  (\bibinfo {year} {2014})}\BibitemShut {NoStop}%
\bibitem [{\citenamefont {Chen}\ \emph {et~al.}(2011)\citenamefont {Chen},
  \citenamefont {Hover}, \citenamefont {Sendelbach}, \citenamefont {Maurer},
  \citenamefont {Merkel}, \citenamefont {Pritchett}, \citenamefont {Wilhelm},\
  and\ \citenamefont {McDermott}}]{Chen2011}%
  \BibitemOpen
  \bibfield  {author} {\bibinfo {author} {\bibfnamefont {Y.-F.}\ \bibnamefont
  {Chen}}, \bibinfo {author} {\bibfnamefont {D.}~\bibnamefont {Hover}},
  \bibinfo {author} {\bibfnamefont {S.}~\bibnamefont {Sendelbach}}, \bibinfo
  {author} {\bibfnamefont {L.}~\bibnamefont {Maurer}}, \bibinfo {author}
  {\bibfnamefont {S.~T.}\ \bibnamefont {Merkel}}, \bibinfo {author}
  {\bibfnamefont {E.~J.}\ \bibnamefont {Pritchett}}, \bibinfo {author}
  {\bibfnamefont {F.~K.}\ \bibnamefont {Wilhelm}}, \ and\ \bibinfo {author}
  {\bibfnamefont {R.}~\bibnamefont {McDermott}},\ }\href {\doibase
  10.1103/PhysRevLett.107.217401} {\bibfield  {journal} {\bibinfo  {journal}
  {Phys. Rev. Lett.}\ }\textbf {\bibinfo {volume} {107}},\ \bibinfo {pages}
  {217401} (\bibinfo {year} {2011})}\BibitemShut {NoStop}%
\bibitem [{\citenamefont {Opremcak}\ \emph {et~al.}(2018)\citenamefont
  {Opremcak}, \citenamefont {Pechenezhskiy}, \citenamefont {Howington},
  \citenamefont {Christensen}, \citenamefont {Beck}, \citenamefont {Leonard},
  \citenamefont {Suttle}, \citenamefont {Wilen}, \citenamefont {Nesterov},
  \citenamefont {Ribeill}, \citenamefont {Thorbeck}, \citenamefont {Schlenker},
  \citenamefont {Vavilov}, \citenamefont {Plourde},\ and\ \citenamefont
  {McDermott}}]{Opremcak2018}%
  \BibitemOpen
  \bibfield  {author} {\bibinfo {author} {\bibfnamefont {A.}~\bibnamefont
  {Opremcak}}, \bibinfo {author} {\bibfnamefont {I.~V.}\ \bibnamefont
  {Pechenezhskiy}}, \bibinfo {author} {\bibfnamefont {C.}~\bibnamefont
  {Howington}}, \bibinfo {author} {\bibfnamefont {B.~G.}\ \bibnamefont
  {Christensen}}, \bibinfo {author} {\bibfnamefont {M.~A.}\ \bibnamefont
  {Beck}}, \bibinfo {author} {\bibfnamefont {E.}~\bibnamefont {Leonard}},
  \bibinfo {author} {\bibfnamefont {J.}~\bibnamefont {Suttle}}, \bibinfo
  {author} {\bibfnamefont {C.}~\bibnamefont {Wilen}}, \bibinfo {author}
  {\bibfnamefont {K.~N.}\ \bibnamefont {Nesterov}}, \bibinfo {author}
  {\bibfnamefont {G.~J.}\ \bibnamefont {Ribeill}}, \bibinfo {author}
  {\bibfnamefont {T.}~\bibnamefont {Thorbeck}}, \bibinfo {author}
  {\bibfnamefont {F.}~\bibnamefont {Schlenker}}, \bibinfo {author}
  {\bibfnamefont {M.~G.}\ \bibnamefont {Vavilov}}, \bibinfo {author}
  {\bibfnamefont {B.~L.~T.}\ \bibnamefont {Plourde}}, \ and\ \bibinfo {author}
  {\bibfnamefont {R.}~\bibnamefont {McDermott}},\ }\href {\doibase
  10.1126/science.aat4625} {\bibfield  {journal} {\bibinfo  {journal}
  {Science}\ }\textbf {\bibinfo {volume} {361}},\ \bibinfo {pages} {1239}
  (\bibinfo {year} {2018})}\BibitemShut {NoStop}%
\bibitem [{\citenamefont {Barone}\ and\ \citenamefont
  {Paternò}(1982)}]{Barone1982}%
  \BibitemOpen
  \bibfield  {author} {\bibinfo {author} {\bibfnamefont {A.}~\bibnamefont
  {Barone}}\ and\ \bibinfo {author} {\bibfnamefont {G.}~\bibnamefont
  {Paternò}},\ }\href {\doibase 10.1002/352760278X} {\emph {\bibinfo {title}
  {Physics and Applications of the Josephson Effect}}}\ (\bibinfo  {publisher}
  {Wiley},\ \bibinfo {year} {1982})\BibitemShut {NoStop}%
\bibitem [{\citenamefont {Wallraff}\ \emph {et~al.}(2005)\citenamefont
  {Wallraff}, \citenamefont {Schuster}, \citenamefont {Blais}, \citenamefont
  {Frunzio}, \citenamefont {Majer}, \citenamefont {Devoret}, \citenamefont
  {Girvin},\ and\ \citenamefont {Schoelkopf}}]{Wallraff2005}%
  \BibitemOpen
  \bibfield  {author} {\bibinfo {author} {\bibfnamefont {A.}~\bibnamefont
  {Wallraff}}, \bibinfo {author} {\bibfnamefont {D.~I.}\ \bibnamefont
  {Schuster}}, \bibinfo {author} {\bibfnamefont {A.}~\bibnamefont {Blais}},
  \bibinfo {author} {\bibfnamefont {L.}~\bibnamefont {Frunzio}}, \bibinfo
  {author} {\bibfnamefont {J.}~\bibnamefont {Majer}}, \bibinfo {author}
  {\bibfnamefont {M.~H.}\ \bibnamefont {Devoret}}, \bibinfo {author}
  {\bibfnamefont {S.~M.}\ \bibnamefont {Girvin}}, \ and\ \bibinfo {author}
  {\bibfnamefont {R.~J.}\ \bibnamefont {Schoelkopf}},\ }\href {\doibase
  10.1103/PhysRevLett.95.060501} {\bibfield  {journal} {\bibinfo  {journal}
  {Phys. Rev. Lett.}\ }\textbf {\bibinfo {volume} {95}},\ \bibinfo {pages}
  {060501} (\bibinfo {year} {2005})}\BibitemShut {NoStop}%
\bibitem [{\citenamefont {Koch}\ \emph {et~al.}(2007)\citenamefont {Koch},
  \citenamefont {Yu}, \citenamefont {Gambetta}, \citenamefont {Houck},
  \citenamefont {Schuster}, \citenamefont {Majer}, \citenamefont {Blais},
  \citenamefont {Devoret}, \citenamefont {Girvin},\ and\ \citenamefont
  {Schoelkopf}}]{Koch2007}%
  \BibitemOpen
  \bibfield  {author} {\bibinfo {author} {\bibfnamefont {J.}~\bibnamefont
  {Koch}}, \bibinfo {author} {\bibfnamefont {T.~M.}\ \bibnamefont {Yu}},
  \bibinfo {author} {\bibfnamefont {J.}~\bibnamefont {Gambetta}}, \bibinfo
  {author} {\bibfnamefont {A.~A.}\ \bibnamefont {Houck}}, \bibinfo {author}
  {\bibfnamefont {D.~I.}\ \bibnamefont {Schuster}}, \bibinfo {author}
  {\bibfnamefont {J.}~\bibnamefont {Majer}}, \bibinfo {author} {\bibfnamefont
  {A.}~\bibnamefont {Blais}}, \bibinfo {author} {\bibfnamefont {M.~H.}\
  \bibnamefont {Devoret}}, \bibinfo {author} {\bibfnamefont {S.~M.}\
  \bibnamefont {Girvin}}, \ and\ \bibinfo {author} {\bibfnamefont {R.~J.}\
  \bibnamefont {Schoelkopf}},\ }\href {\doibase 10.1103/PhysRevA.76.042319}
  {\bibfield  {journal} {\bibinfo  {journal} {Phys. Rev. A}\ }\textbf {\bibinfo
  {volume} {76}},\ \bibinfo {pages} {042319} (\bibinfo {year}
  {2007})}\BibitemShut {NoStop}%
\bibitem [{\citenamefont {Barends}\ \emph {et~al.}(2013)\citenamefont
  {Barends}, \citenamefont {Kelly}, \citenamefont {Megrant}, \citenamefont
  {Sank}, \citenamefont {Jeffrey}, \citenamefont {Chen}, \citenamefont {Yin},
  \citenamefont {Chiaro}, \citenamefont {Mutus}, \citenamefont {Neill},
  \citenamefont {O'Malley}, \citenamefont {Roushan}, \citenamefont {Wenner},
  \citenamefont {White}, \citenamefont {Cleland},\ and\ \citenamefont
  {Martinis}}]{Barends2013}%
  \BibitemOpen
  \bibfield  {author} {\bibinfo {author} {\bibfnamefont {R.}~\bibnamefont
  {Barends}}, \bibinfo {author} {\bibfnamefont {J.}~\bibnamefont {Kelly}},
  \bibinfo {author} {\bibfnamefont {A.}~\bibnamefont {Megrant}}, \bibinfo
  {author} {\bibfnamefont {D.}~\bibnamefont {Sank}}, \bibinfo {author}
  {\bibfnamefont {E.}~\bibnamefont {Jeffrey}}, \bibinfo {author} {\bibfnamefont
  {Y.}~\bibnamefont {Chen}}, \bibinfo {author} {\bibfnamefont {Y.}~\bibnamefont
  {Yin}}, \bibinfo {author} {\bibfnamefont {B.}~\bibnamefont {Chiaro}},
  \bibinfo {author} {\bibfnamefont {J.}~\bibnamefont {Mutus}}, \bibinfo
  {author} {\bibfnamefont {C.}~\bibnamefont {Neill}}, \bibinfo {author}
  {\bibfnamefont {P.}~\bibnamefont {O'Malley}}, \bibinfo {author}
  {\bibfnamefont {P.}~\bibnamefont {Roushan}}, \bibinfo {author} {\bibfnamefont
  {J.}~\bibnamefont {Wenner}}, \bibinfo {author} {\bibfnamefont {T.~C.}\
  \bibnamefont {White}}, \bibinfo {author} {\bibfnamefont {A.~N.}\ \bibnamefont
  {Cleland}}, \ and\ \bibinfo {author} {\bibfnamefont {J.~M.}\ \bibnamefont
  {Martinis}},\ }\href {\doibase 10.1103/PhysRevLett.111.080502} {\bibfield
  {journal} {\bibinfo  {journal} {Phys. Rev. Lett.}\ }\textbf {\bibinfo
  {volume} {111}},\ \bibinfo {pages} {080502} (\bibinfo {year}
  {2013})}\BibitemShut {NoStop}%
\bibitem [{\citenamefont {Houck}\ \emph {et~al.}(2008)\citenamefont {Houck},
  \citenamefont {Schreier}, \citenamefont {Johnson}, \citenamefont {Chow},
  \citenamefont {Koch}, \citenamefont {Gambetta}, \citenamefont {Schuster},
  \citenamefont {Frunzio}, \citenamefont {Devoret}, \citenamefont {Girvin},\
  and\ \citenamefont {Schoelkopf}}]{Houck2008}%
  \BibitemOpen
  \bibfield  {author} {\bibinfo {author} {\bibfnamefont {A.~A.}\ \bibnamefont
  {Houck}}, \bibinfo {author} {\bibfnamefont {J.~A.}\ \bibnamefont {Schreier}},
  \bibinfo {author} {\bibfnamefont {B.~R.}\ \bibnamefont {Johnson}}, \bibinfo
  {author} {\bibfnamefont {J.~M.}\ \bibnamefont {Chow}}, \bibinfo {author}
  {\bibfnamefont {J.}~\bibnamefont {Koch}}, \bibinfo {author} {\bibfnamefont
  {J.~M.}\ \bibnamefont {Gambetta}}, \bibinfo {author} {\bibfnamefont {D.~I.}\
  \bibnamefont {Schuster}}, \bibinfo {author} {\bibfnamefont {L.}~\bibnamefont
  {Frunzio}}, \bibinfo {author} {\bibfnamefont {M.~H.}\ \bibnamefont
  {Devoret}}, \bibinfo {author} {\bibfnamefont {S.~M.}\ \bibnamefont {Girvin}},
  \ and\ \bibinfo {author} {\bibfnamefont {R.~J.}\ \bibnamefont {Schoelkopf}},\
  }\href {\doibase 10.1103/PhysRevLett.101.080502} {\bibfield  {journal}
  {\bibinfo  {journal} {Phys. Rev. Lett.}\ }\textbf {\bibinfo {volume} {101}},\
  \bibinfo {pages} {080502} (\bibinfo {year} {2008})}\BibitemShut {NoStop}%
\bibitem [{\citenamefont {Motzoi}\ \emph {et~al.}(2009)\citenamefont {Motzoi},
  \citenamefont {Gambetta}, \citenamefont {Rebentrost},\ and\ \citenamefont
  {Wilhelm}}]{Motzoi2009}%
  \BibitemOpen
  \bibfield  {author} {\bibinfo {author} {\bibfnamefont {F.}~\bibnamefont
  {Motzoi}}, \bibinfo {author} {\bibfnamefont {J.~M.}\ \bibnamefont
  {Gambetta}}, \bibinfo {author} {\bibfnamefont {P.}~\bibnamefont
  {Rebentrost}}, \ and\ \bibinfo {author} {\bibfnamefont {F.~K.}\ \bibnamefont
  {Wilhelm}},\ }\href {\doibase 10.1103/PhysRevLett.103.110501} {\bibfield
  {journal} {\bibinfo  {journal} {Phys. Rev. Lett.}\ }\textbf {\bibinfo
  {volume} {103}},\ \bibinfo {pages} {110501} (\bibinfo {year}
  {2009})}\BibitemShut {NoStop}%
\bibitem [{\citenamefont {Lucero}\ \emph {et~al.}(2010)\citenamefont {Lucero},
  \citenamefont {Kelly}, \citenamefont {Bialczak}, \citenamefont {Lenander},
  \citenamefont {Mariantoni}, \citenamefont {Neeley}, \citenamefont
  {O'Connell}, \citenamefont {Sank}, \citenamefont {Wang}, \citenamefont
  {Weides}, \citenamefont {Wenner}, \citenamefont {Yamamoto}, \citenamefont
  {Cleland},\ and\ \citenamefont {Martinis}}]{Lucero2010}%
  \BibitemOpen
  \bibfield  {author} {\bibinfo {author} {\bibfnamefont {E.}~\bibnamefont
  {Lucero}}, \bibinfo {author} {\bibfnamefont {J.}~\bibnamefont {Kelly}},
  \bibinfo {author} {\bibfnamefont {R.~C.}\ \bibnamefont {Bialczak}}, \bibinfo
  {author} {\bibfnamefont {M.}~\bibnamefont {Lenander}}, \bibinfo {author}
  {\bibfnamefont {M.}~\bibnamefont {Mariantoni}}, \bibinfo {author}
  {\bibfnamefont {M.}~\bibnamefont {Neeley}}, \bibinfo {author} {\bibfnamefont
  {A.~D.}\ \bibnamefont {O'Connell}}, \bibinfo {author} {\bibfnamefont
  {D.}~\bibnamefont {Sank}}, \bibinfo {author} {\bibfnamefont {H.}~\bibnamefont
  {Wang}}, \bibinfo {author} {\bibfnamefont {M.}~\bibnamefont {Weides}},
  \bibinfo {author} {\bibfnamefont {J.}~\bibnamefont {Wenner}}, \bibinfo
  {author} {\bibfnamefont {T.}~\bibnamefont {Yamamoto}}, \bibinfo {author}
  {\bibfnamefont {A.~N.}\ \bibnamefont {Cleland}}, \ and\ \bibinfo {author}
  {\bibfnamefont {J.~M.}\ \bibnamefont {Martinis}},\ }\href {\doibase
  10.1103/PhysRevA.82.042339} {\bibfield  {journal} {\bibinfo  {journal} {Phys.
  Rev. A}\ }\textbf {\bibinfo {volume} {82}},\ \bibinfo {pages} {042339}
  (\bibinfo {year} {2010})}\BibitemShut {NoStop}%
\bibitem [{\citenamefont {Chen}\ \emph {et~al.}(2016)\citenamefont {Chen},
  \citenamefont {Kelly}, \citenamefont {Quintana}, \citenamefont {Barends},
  \citenamefont {Campbell}, \citenamefont {Chen}, \citenamefont {Chiaro},
  \citenamefont {Dunsworth}, \citenamefont {Fowler}, \citenamefont {Lucero},
  \citenamefont {Jeffrey}, \citenamefont {Megrant}, \citenamefont {Mutus},
  \citenamefont {Neeley}, \citenamefont {Neill}, \citenamefont {O'Malley},
  \citenamefont {Roushan}, \citenamefont {Sank}, \citenamefont {Vainsencher},
  \citenamefont {Wenner}, \citenamefont {White}, \citenamefont {Korotkov},\
  and\ \citenamefont {Martinis}}]{Chen2016}%
  \BibitemOpen
  \bibfield  {author} {\bibinfo {author} {\bibfnamefont {Z.}~\bibnamefont
  {Chen}}, \bibinfo {author} {\bibfnamefont {J.}~\bibnamefont {Kelly}},
  \bibinfo {author} {\bibfnamefont {C.}~\bibnamefont {Quintana}}, \bibinfo
  {author} {\bibfnamefont {R.}~\bibnamefont {Barends}}, \bibinfo {author}
  {\bibfnamefont {B.}~\bibnamefont {Campbell}}, \bibinfo {author}
  {\bibfnamefont {Y.}~\bibnamefont {Chen}}, \bibinfo {author} {\bibfnamefont
  {B.}~\bibnamefont {Chiaro}}, \bibinfo {author} {\bibfnamefont
  {A.}~\bibnamefont {Dunsworth}}, \bibinfo {author} {\bibfnamefont {A.~G.}\
  \bibnamefont {Fowler}}, \bibinfo {author} {\bibfnamefont {E.}~\bibnamefont
  {Lucero}}, \bibinfo {author} {\bibfnamefont {E.}~\bibnamefont {Jeffrey}},
  \bibinfo {author} {\bibfnamefont {A.}~\bibnamefont {Megrant}}, \bibinfo
  {author} {\bibfnamefont {J.}~\bibnamefont {Mutus}}, \bibinfo {author}
  {\bibfnamefont {M.}~\bibnamefont {Neeley}}, \bibinfo {author} {\bibfnamefont
  {C.}~\bibnamefont {Neill}}, \bibinfo {author} {\bibfnamefont {P.~J.~J.}\
  \bibnamefont {O'Malley}}, \bibinfo {author} {\bibfnamefont {P.}~\bibnamefont
  {Roushan}}, \bibinfo {author} {\bibfnamefont {D.}~\bibnamefont {Sank}},
  \bibinfo {author} {\bibfnamefont {A.}~\bibnamefont {Vainsencher}}, \bibinfo
  {author} {\bibfnamefont {J.}~\bibnamefont {Wenner}}, \bibinfo {author}
  {\bibfnamefont {T.~C.}\ \bibnamefont {White}}, \bibinfo {author}
  {\bibfnamefont {A.~N.}\ \bibnamefont {Korotkov}}, \ and\ \bibinfo {author}
  {\bibfnamefont {J.~M.}\ \bibnamefont {Martinis}},\ }\href {\doibase
  10.1103/PhysRevLett.116.020501} {\bibfield  {journal} {\bibinfo  {journal}
  {Phys. Rev. Lett.}\ }\textbf {\bibinfo {volume} {116}},\ \bibinfo {pages}
  {020501} (\bibinfo {year} {2016})}\BibitemShut {NoStop}%
\bibitem [{\citenamefont {Hofheinz}\ \emph {et~al.}(2008)\citenamefont
  {Hofheinz}, \citenamefont {Weig}, \citenamefont {Ansmann}, \citenamefont
  {Bialczak}, \citenamefont {Lucero}, \citenamefont {Neeley}, \citenamefont
  {O'Connell}, \citenamefont {Wang}, \citenamefont {Martinis},\ and\
  \citenamefont {Cleland}}]{Hofheinz2008}%
  \BibitemOpen
  \bibfield  {author} {\bibinfo {author} {\bibfnamefont {M.}~\bibnamefont
  {Hofheinz}}, \bibinfo {author} {\bibfnamefont {E.~M.}\ \bibnamefont {Weig}},
  \bibinfo {author} {\bibfnamefont {M.}~\bibnamefont {Ansmann}}, \bibinfo
  {author} {\bibfnamefont {R.~C.}\ \bibnamefont {Bialczak}}, \bibinfo {author}
  {\bibfnamefont {E.}~\bibnamefont {Lucero}}, \bibinfo {author} {\bibfnamefont
  {M.}~\bibnamefont {Neeley}}, \bibinfo {author} {\bibfnamefont {A.~D.}\
  \bibnamefont {O'Connell}}, \bibinfo {author} {\bibfnamefont {H.}~\bibnamefont
  {Wang}}, \bibinfo {author} {\bibfnamefont {J.~M.}\ \bibnamefont {Martinis}},
  \ and\ \bibinfo {author} {\bibfnamefont {A.~N.}\ \bibnamefont {Cleland}},\
  }\href {\doibase 10.1038/nature07136} {\bibfield  {journal} {\bibinfo
  {journal} {Nature}\ }\textbf {\bibinfo {volume} {454}},\ \bibinfo {pages}
  {310} (\bibinfo {year} {2008})}\BibitemShut {NoStop}%
\bibitem [{\citenamefont {Cooper}\ \emph {et~al.}(2004)\citenamefont {Cooper},
  \citenamefont {Steffen}, \citenamefont {McDermott}, \citenamefont {Simmonds},
  \citenamefont {Oh}, \citenamefont {Hite}, \citenamefont {Pappas},\ and\
  \citenamefont {Martinis}}]{Cooper2004}%
  \BibitemOpen
  \bibfield  {author} {\bibinfo {author} {\bibfnamefont {K.~B.}\ \bibnamefont
  {Cooper}}, \bibinfo {author} {\bibfnamefont {M.}~\bibnamefont {Steffen}},
  \bibinfo {author} {\bibfnamefont {R.}~\bibnamefont {McDermott}}, \bibinfo
  {author} {\bibfnamefont {R.~W.}\ \bibnamefont {Simmonds}}, \bibinfo {author}
  {\bibfnamefont {S.}~\bibnamefont {Oh}}, \bibinfo {author} {\bibfnamefont
  {D.~A.}\ \bibnamefont {Hite}}, \bibinfo {author} {\bibfnamefont {D.~P.}\
  \bibnamefont {Pappas}}, \ and\ \bibinfo {author} {\bibfnamefont {J.~M.}\
  \bibnamefont {Martinis}},\ }\href {\doibase 10.1103/PhysRevLett.93.180401}
  {\bibfield  {journal} {\bibinfo  {journal} {Phys. Rev. Lett.}\ }\textbf
  {\bibinfo {volume} {93}},\ \bibinfo {pages} {180401} (\bibinfo {year}
  {2004})}\BibitemShut {NoStop}%
\bibitem [{\citenamefont {Khezri}\ \emph {et~al.}(2016)\citenamefont {Khezri},
  \citenamefont {Mlinar}, \citenamefont {Dressel},\ and\ \citenamefont
  {Korotkov}}]{Khezri2016}%
  \BibitemOpen
  \bibfield  {author} {\bibinfo {author} {\bibfnamefont {M.}~\bibnamefont
  {Khezri}}, \bibinfo {author} {\bibfnamefont {E.}~\bibnamefont {Mlinar}},
  \bibinfo {author} {\bibfnamefont {J.}~\bibnamefont {Dressel}}, \ and\
  \bibinfo {author} {\bibfnamefont {A.~N.}\ \bibnamefont {Korotkov}},\ }\href
  {\doibase 10.1103/PhysRevA.94.012347} {\bibfield  {journal} {\bibinfo
  {journal} {Phys. Rev. A}\ }\textbf {\bibinfo {volume} {94}},\ \bibinfo
  {pages} {012347} (\bibinfo {year} {2016})}\BibitemShut {NoStop}%
\bibitem [{\citenamefont {Khezri}\ and\ \citenamefont
  {Korotkov}(2017)}]{Khezri2017}%
  \BibitemOpen
  \bibfield  {author} {\bibinfo {author} {\bibfnamefont {M.}~\bibnamefont
  {Khezri}}\ and\ \bibinfo {author} {\bibfnamefont {A.~N.}\ \bibnamefont
  {Korotkov}},\ }\href {\doibase 10.1103/PhysRevA.96.043839} {\bibfield
  {journal} {\bibinfo  {journal} {Phys. Rev. A}\ }\textbf {\bibinfo {volume}
  {96}},\ \bibinfo {pages} {043839} (\bibinfo {year} {2017})}\BibitemShut
  {NoStop}%
\bibitem [{\citenamefont {Gambetta}\ \emph {et~al.}(2007)\citenamefont
  {Gambetta}, \citenamefont {Braff}, \citenamefont {Wallraff}, \citenamefont
  {Girvin},\ and\ \citenamefont {Schoelkopf}}]{Gambetta2007}%
  \BibitemOpen
  \bibfield  {author} {\bibinfo {author} {\bibfnamefont {J.}~\bibnamefont
  {Gambetta}}, \bibinfo {author} {\bibfnamefont {W.~A.}\ \bibnamefont {Braff}},
  \bibinfo {author} {\bibfnamefont {A.}~\bibnamefont {Wallraff}}, \bibinfo
  {author} {\bibfnamefont {S.~M.}\ \bibnamefont {Girvin}}, \ and\ \bibinfo
  {author} {\bibfnamefont {R.~J.}\ \bibnamefont {Schoelkopf}},\ }\href
  {\doibase 10.1103/PhysRevA.76.012325} {\bibfield  {journal} {\bibinfo
  {journal} {Phys. Rev. A}\ }\textbf {\bibinfo {volume} {76}},\ \bibinfo
  {pages} {012325} (\bibinfo {year} {2007})}\BibitemShut {NoStop}%
\bibitem [{\citenamefont {Magesan}\ \emph {et~al.}(2012)\citenamefont
  {Magesan}, \citenamefont {Gambetta}, \citenamefont {Johnson}, \citenamefont
  {Ryan}, \citenamefont {Chow}, \citenamefont {Merkel}, \citenamefont
  {da~Silva}, \citenamefont {Keefe}, \citenamefont {Rothwell}, \citenamefont
  {Ohki}, \citenamefont {Ketchen},\ and\ \citenamefont
  {Steffen}}]{Magesan2012}%
  \BibitemOpen
  \bibfield  {author} {\bibinfo {author} {\bibfnamefont {E.}~\bibnamefont
  {Magesan}}, \bibinfo {author} {\bibfnamefont {J.~M.}\ \bibnamefont
  {Gambetta}}, \bibinfo {author} {\bibfnamefont {B.~R.}\ \bibnamefont
  {Johnson}}, \bibinfo {author} {\bibfnamefont {C.~A.}\ \bibnamefont {Ryan}},
  \bibinfo {author} {\bibfnamefont {J.~M.}\ \bibnamefont {Chow}}, \bibinfo
  {author} {\bibfnamefont {S.~T.}\ \bibnamefont {Merkel}}, \bibinfo {author}
  {\bibfnamefont {M.~P.}\ \bibnamefont {da~Silva}}, \bibinfo {author}
  {\bibfnamefont {G.~A.}\ \bibnamefont {Keefe}}, \bibinfo {author}
  {\bibfnamefont {M.~B.}\ \bibnamefont {Rothwell}}, \bibinfo {author}
  {\bibfnamefont {T.~A.}\ \bibnamefont {Ohki}}, \bibinfo {author}
  {\bibfnamefont {M.~B.}\ \bibnamefont {Ketchen}}, \ and\ \bibinfo {author}
  {\bibfnamefont {M.}~\bibnamefont {Steffen}},\ }\href {\doibase
  10.1103/PhysRevLett.109.080505} {\bibfield  {journal} {\bibinfo  {journal}
  {Phys. Rev. Lett.}\ }\textbf {\bibinfo {volume} {109}},\ \bibinfo {pages}
  {080505} (\bibinfo {year} {2012})}\BibitemShut {NoStop}%
\bibitem [{\citenamefont {McDermott}\ \emph {et~al.}(2005)\citenamefont
  {McDermott}, \citenamefont {Simmonds}, \citenamefont {Steffen}, \citenamefont
  {Cooper}, \citenamefont {Cicak}, \citenamefont {Osborn}, \citenamefont {Oh},
  \citenamefont {Pappas},\ and\ \citenamefont {Martinis}}]{McDermott2005}%
  \BibitemOpen
  \bibfield  {author} {\bibinfo {author} {\bibfnamefont {R.}~\bibnamefont
  {McDermott}}, \bibinfo {author} {\bibfnamefont {R.~W.}\ \bibnamefont
  {Simmonds}}, \bibinfo {author} {\bibfnamefont {M.}~\bibnamefont {Steffen}},
  \bibinfo {author} {\bibfnamefont {K.~B.}\ \bibnamefont {Cooper}}, \bibinfo
  {author} {\bibfnamefont {K.}~\bibnamefont {Cicak}}, \bibinfo {author}
  {\bibfnamefont {K.~D.}\ \bibnamefont {Osborn}}, \bibinfo {author}
  {\bibfnamefont {S.}~\bibnamefont {Oh}}, \bibinfo {author} {\bibfnamefont
  {D.~P.}\ \bibnamefont {Pappas}}, \ and\ \bibinfo {author} {\bibfnamefont
  {J.~M.}\ \bibnamefont {Martinis}},\ }\href {\doibase 10.1126/science.1107572}
  {\bibfield  {journal} {\bibinfo  {journal} {Science}\ }\textbf {\bibinfo
  {volume} {307}},\ \bibinfo {pages} {1299} (\bibinfo {year}
  {2005})}\BibitemShut {NoStop}%
\bibitem [{\citenamefont {Martinis}\ \emph {et~al.}(2003)\citenamefont
  {Martinis}, \citenamefont {Nam}, \citenamefont {Aumentado}, \citenamefont
  {Lang},\ and\ \citenamefont {Urbina}}]{Martinis2003}%
  \BibitemOpen
  \bibfield  {author} {\bibinfo {author} {\bibfnamefont {J.~M.}\ \bibnamefont
  {Martinis}}, \bibinfo {author} {\bibfnamefont {S.}~\bibnamefont {Nam}},
  \bibinfo {author} {\bibfnamefont {J.}~\bibnamefont {Aumentado}}, \bibinfo
  {author} {\bibfnamefont {K.~M.}\ \bibnamefont {Lang}}, \ and\ \bibinfo
  {author} {\bibfnamefont {C.}~\bibnamefont {Urbina}},\ }\href {\doibase
  10.1103/PhysRevB.67.094510} {\bibfield  {journal} {\bibinfo  {journal} {Phys.
  Rev. B}\ }\textbf {\bibinfo {volume} {67}},\ \bibinfo {pages} {094510}
  (\bibinfo {year} {2003})}\BibitemShut {NoStop}%
\bibitem [{\citenamefont {Bylander}\ \emph {et~al.}(2011)\citenamefont
  {Bylander}, \citenamefont {Gustavsson}, \citenamefont {Yan}, \citenamefont
  {Yoshihara}, \citenamefont {Harrabi}, \citenamefont {Fitch}, \citenamefont
  {Cory}, \citenamefont {Nakamura}, \citenamefont {Tsai},\ and\ \citenamefont
  {Oliver}}]{Bylander2011}%
  \BibitemOpen
  \bibfield  {author} {\bibinfo {author} {\bibfnamefont {J.}~\bibnamefont
  {Bylander}}, \bibinfo {author} {\bibfnamefont {S.}~\bibnamefont
  {Gustavsson}}, \bibinfo {author} {\bibfnamefont {F.}~\bibnamefont {Yan}},
  \bibinfo {author} {\bibfnamefont {F.}~\bibnamefont {Yoshihara}}, \bibinfo
  {author} {\bibfnamefont {K.}~\bibnamefont {Harrabi}}, \bibinfo {author}
  {\bibfnamefont {G.}~\bibnamefont {Fitch}}, \bibinfo {author} {\bibfnamefont
  {D.~G.}\ \bibnamefont {Cory}}, \bibinfo {author} {\bibfnamefont
  {Y.}~\bibnamefont {Nakamura}}, \bibinfo {author} {\bibfnamefont {J.-S.}\
  \bibnamefont {Tsai}}, \ and\ \bibinfo {author} {\bibfnamefont {W.~D.}\
  \bibnamefont {Oliver}},\ }\href {\doibase 10.1038/nphys1994} {\bibfield
  {journal} {\bibinfo  {journal} {Nat. Phys.}\ }\textbf {\bibinfo {volume}
  {7}},\ \bibinfo {pages} {565} (\bibinfo {year} {2011})}\BibitemShut {NoStop}%
\bibitem [{\citenamefont {Hutchings}\ \emph {et~al.}(2017)\citenamefont
  {Hutchings}, \citenamefont {Hertzberg}, \citenamefont {Liu}, \citenamefont
  {Bronn}, \citenamefont {Keefe}, \citenamefont {Brink}, \citenamefont {Chow},\
  and\ \citenamefont {Plourde}}]{Hutchings2017}%
  \BibitemOpen
  \bibfield  {author} {\bibinfo {author} {\bibfnamefont {M.~D.}\ \bibnamefont
  {Hutchings}}, \bibinfo {author} {\bibfnamefont {J.~B.}\ \bibnamefont
  {Hertzberg}}, \bibinfo {author} {\bibfnamefont {Y.}~\bibnamefont {Liu}},
  \bibinfo {author} {\bibfnamefont {N.~T.}\ \bibnamefont {Bronn}}, \bibinfo
  {author} {\bibfnamefont {G.~A.}\ \bibnamefont {Keefe}}, \bibinfo {author}
  {\bibfnamefont {M.}~\bibnamefont {Brink}}, \bibinfo {author} {\bibfnamefont
  {J.~M.}\ \bibnamefont {Chow}}, \ and\ \bibinfo {author} {\bibfnamefont
  {B.~L.~T.}\ \bibnamefont {Plourde}},\ }\href {\doibase
  10.1103/PhysRevApplied.8.044003} {\bibfield  {journal} {\bibinfo  {journal}
  {Phys. Rev. Applied}\ }\textbf {\bibinfo {volume} {8}},\ \bibinfo {pages}
  {044003} (\bibinfo {year} {2017})}\BibitemShut {NoStop}%
\bibitem [{Note1()}]{Note1}%
  \BibitemOpen
  \bibinfo {note} {\protect \leavevmode {\protect \color {black}The spin-echo
  decay envelope obtained from modeling photon shot noise dephasing using our
  circuit parameters is well described by a Gaussian function of the idle gate
  time}}\BibitemShut {NoStop}%
\bibitem [{\citenamefont {Sears}\ \emph {et~al.}(2012)\citenamefont {Sears},
  \citenamefont {Petrenko}, \citenamefont {Catelani}, \citenamefont {Sun},
  \citenamefont {Paik}, \citenamefont {Kirchmair}, \citenamefont {Frunzio},
  \citenamefont {Glazman}, \citenamefont {Girvin},\ and\ \citenamefont
  {Schoelkopf}}]{Sears2012}%
  \BibitemOpen
  \bibfield  {author} {\bibinfo {author} {\bibfnamefont {A.~P.}\ \bibnamefont
  {Sears}}, \bibinfo {author} {\bibfnamefont {A.}~\bibnamefont {Petrenko}},
  \bibinfo {author} {\bibfnamefont {G.}~\bibnamefont {Catelani}}, \bibinfo
  {author} {\bibfnamefont {L.}~\bibnamefont {Sun}}, \bibinfo {author}
  {\bibfnamefont {H.}~\bibnamefont {Paik}}, \bibinfo {author} {\bibfnamefont
  {G.}~\bibnamefont {Kirchmair}}, \bibinfo {author} {\bibfnamefont
  {L.}~\bibnamefont {Frunzio}}, \bibinfo {author} {\bibfnamefont {L.~I.}\
  \bibnamefont {Glazman}}, \bibinfo {author} {\bibfnamefont {S.~M.}\
  \bibnamefont {Girvin}}, \ and\ \bibinfo {author} {\bibfnamefont {R.~J.}\
  \bibnamefont {Schoelkopf}},\ }\href {\doibase 10.1103/PhysRevB.86.180504}
  {\bibfield  {journal} {\bibinfo  {journal} {Phys. Rev. B}\ }\textbf {\bibinfo
  {volume} {86}},\ \bibinfo {pages} {180504} (\bibinfo {year}
  {2012})}\BibitemShut {NoStop}%
\bibitem [{\citenamefont {Yan}\ \emph {et~al.}(2016)\citenamefont {Yan},
  \citenamefont {Gustavsson}, \citenamefont {Kamal}, \citenamefont {Birenbaum},
  \citenamefont {Sears}, \citenamefont {Hover}, \citenamefont {Gudmundsen},
  \citenamefont {Rosenberg}, \citenamefont {Samach}, \citenamefont {Weber},
  \citenamefont {Yoder}, \citenamefont {Orlando}, \citenamefont {Clarke},
  \citenamefont {Kerman},\ and\ \citenamefont {Oliver}}]{Yan2016}%
  \BibitemOpen
  \bibfield  {author} {\bibinfo {author} {\bibfnamefont {F.}~\bibnamefont
  {Yan}}, \bibinfo {author} {\bibfnamefont {S.}~\bibnamefont {Gustavsson}},
  \bibinfo {author} {\bibfnamefont {A.}~\bibnamefont {Kamal}}, \bibinfo
  {author} {\bibfnamefont {J.}~\bibnamefont {Birenbaum}}, \bibinfo {author}
  {\bibfnamefont {A.~P.}\ \bibnamefont {Sears}}, \bibinfo {author}
  {\bibfnamefont {D.}~\bibnamefont {Hover}}, \bibinfo {author} {\bibfnamefont
  {T.~J.}\ \bibnamefont {Gudmundsen}}, \bibinfo {author} {\bibfnamefont
  {D.}~\bibnamefont {Rosenberg}}, \bibinfo {author} {\bibfnamefont
  {G.}~\bibnamefont {Samach}}, \bibinfo {author} {\bibfnamefont
  {S.}~\bibnamefont {Weber}}, \bibinfo {author} {\bibfnamefont {J.~L.}\
  \bibnamefont {Yoder}}, \bibinfo {author} {\bibfnamefont {T.~P.}\ \bibnamefont
  {Orlando}}, \bibinfo {author} {\bibfnamefont {J.}~\bibnamefont {Clarke}},
  \bibinfo {author} {\bibfnamefont {A.~J.}\ \bibnamefont {Kerman}}, \ and\
  \bibinfo {author} {\bibfnamefont {W.~D.}\ \bibnamefont {Oliver}},\ }\href
  {\doibase 10.1038/ncomms12964} {\bibfield  {journal} {\bibinfo  {journal}
  {Nat. Commun.}\ }\textbf {\bibinfo {volume} {7}},\ \bibinfo {pages} {12964}
  (\bibinfo {year} {2016})}\BibitemShut {NoStop}%
\bibitem [{\citenamefont {Yan}\ \emph {et~al.}(2018)\citenamefont {Yan},
  \citenamefont {Campbell}, \citenamefont {Krantz}, \citenamefont {Kjaergaard},
  \citenamefont {Kim}, \citenamefont {Yoder}, \citenamefont {Hover},
  \citenamefont {Sears}, \citenamefont {Kerman}, \citenamefont {Orlando},
  \citenamefont {Gustavsson},\ and\ \citenamefont {Oliver}}]{Yan2018}%
  \BibitemOpen
  \bibfield  {author} {\bibinfo {author} {\bibfnamefont {F.}~\bibnamefont
  {Yan}}, \bibinfo {author} {\bibfnamefont {D.}~\bibnamefont {Campbell}},
  \bibinfo {author} {\bibfnamefont {P.}~\bibnamefont {Krantz}}, \bibinfo
  {author} {\bibfnamefont {M.}~\bibnamefont {Kjaergaard}}, \bibinfo {author}
  {\bibfnamefont {D.}~\bibnamefont {Kim}}, \bibinfo {author} {\bibfnamefont
  {J.~L.}\ \bibnamefont {Yoder}}, \bibinfo {author} {\bibfnamefont
  {D.}~\bibnamefont {Hover}}, \bibinfo {author} {\bibfnamefont
  {A.}~\bibnamefont {Sears}}, \bibinfo {author} {\bibfnamefont {A.~J.}\
  \bibnamefont {Kerman}}, \bibinfo {author} {\bibfnamefont {T.~P.}\
  \bibnamefont {Orlando}}, \bibinfo {author} {\bibfnamefont {S.}~\bibnamefont
  {Gustavsson}}, \ and\ \bibinfo {author} {\bibfnamefont {W.~D.}\ \bibnamefont
  {Oliver}},\ }\href {\doibase 10.1103/PhysRevLett.120.260504} {\bibfield
  {journal} {\bibinfo  {journal} {Phys. Rev. Lett.}\ }\textbf {\bibinfo
  {volume} {120}},\ \bibinfo {pages} {260504} (\bibinfo {year}
  {2018})}\BibitemShut {NoStop}%
\bibitem [{\citenamefont {Andersen}\ \emph {et~al.}(2019)\citenamefont
  {Andersen}, \citenamefont {Remm}, \citenamefont {Lazar}, \citenamefont
  {Krinner}, \citenamefont {Heinsoo}, \citenamefont {Besse}, \citenamefont
  {Gabureac}, \citenamefont {Wallraff},\ and\ \citenamefont
  {Eichler}}]{Andersen2019}%
  \BibitemOpen
  \bibfield  {author} {\bibinfo {author} {\bibfnamefont {C.~K.}\ \bibnamefont
  {Andersen}}, \bibinfo {author} {\bibfnamefont {A.}~\bibnamefont {Remm}},
  \bibinfo {author} {\bibfnamefont {S.}~\bibnamefont {Lazar}}, \bibinfo
  {author} {\bibfnamefont {S.}~\bibnamefont {Krinner}}, \bibinfo {author}
  {\bibfnamefont {J.}~\bibnamefont {Heinsoo}}, \bibinfo {author} {\bibfnamefont
  {J.-C.}\ \bibnamefont {Besse}}, \bibinfo {author} {\bibfnamefont
  {M.}~\bibnamefont {Gabureac}}, \bibinfo {author} {\bibfnamefont
  {A.}~\bibnamefont {Wallraff}}, \ and\ \bibinfo {author} {\bibfnamefont
  {C.}~\bibnamefont {Eichler}},\ }\href {\doibase 10.1038/s41534-019-0185-4}
  {\bibfield  {journal} {\bibinfo  {journal} {npj Quantum Inf.}\ }\textbf
  {\bibinfo {volume} {5}},\ \bibinfo {pages} {69} (\bibinfo {year}
  {2019})}\BibitemShut {NoStop}%
\bibitem [{\citenamefont {Andersen}\ \emph {et~al.}(2020)\citenamefont
  {Andersen}, \citenamefont {Remm}, \citenamefont {Lazar}, \citenamefont
  {Krinner}, \citenamefont {Lacroix}, \citenamefont {Norris}, \citenamefont
  {Gabureac}, \citenamefont {Eichler},\ and\ \citenamefont
  {Wallraff}}]{Andersen2020}%
  \BibitemOpen
  \bibfield  {author} {\bibinfo {author} {\bibfnamefont {C.~K.}\ \bibnamefont
  {Andersen}}, \bibinfo {author} {\bibfnamefont {A.}~\bibnamefont {Remm}},
  \bibinfo {author} {\bibfnamefont {S.}~\bibnamefont {Lazar}}, \bibinfo
  {author} {\bibfnamefont {S.}~\bibnamefont {Krinner}}, \bibinfo {author}
  {\bibfnamefont {N.}~\bibnamefont {Lacroix}}, \bibinfo {author} {\bibfnamefont
  {G.~J.}\ \bibnamefont {Norris}}, \bibinfo {author} {\bibfnamefont
  {M.}~\bibnamefont {Gabureac}}, \bibinfo {author} {\bibfnamefont
  {C.}~\bibnamefont {Eichler}}, \ and\ \bibinfo {author} {\bibfnamefont
  {A.}~\bibnamefont {Wallraff}},\ }\href {\doibase 10.1038/s41567-020-0920-y}
  {\bibfield  {journal} {\bibinfo  {journal} {Nature Physics}\ }\textbf
  {\bibinfo {volume} {16}},\ \bibinfo {pages} {875} (\bibinfo {year}
  {2020})}\BibitemShut {NoStop}%
\bibitem [{\citenamefont {Lenander}\ \emph {et~al.}(2011)\citenamefont
  {Lenander}, \citenamefont {Wang}, \citenamefont {Bialczak}, \citenamefont
  {Lucero}, \citenamefont {Mariantoni}, \citenamefont {Neeley}, \citenamefont
  {O'Connell}, \citenamefont {Sank}, \citenamefont {Weides}, \citenamefont
  {Wenner}, \citenamefont {Yamamoto}, \citenamefont {Yin}, \citenamefont
  {Zhao}, \citenamefont {Cleland},\ and\ \citenamefont
  {Martinis}}]{Lenander2011}%
  \BibitemOpen
  \bibfield  {author} {\bibinfo {author} {\bibfnamefont {M.}~\bibnamefont
  {Lenander}}, \bibinfo {author} {\bibfnamefont {H.}~\bibnamefont {Wang}},
  \bibinfo {author} {\bibfnamefont {R.~C.}\ \bibnamefont {Bialczak}}, \bibinfo
  {author} {\bibfnamefont {E.}~\bibnamefont {Lucero}}, \bibinfo {author}
  {\bibfnamefont {M.}~\bibnamefont {Mariantoni}}, \bibinfo {author}
  {\bibfnamefont {M.}~\bibnamefont {Neeley}}, \bibinfo {author} {\bibfnamefont
  {A.~D.}\ \bibnamefont {O'Connell}}, \bibinfo {author} {\bibfnamefont
  {D.}~\bibnamefont {Sank}}, \bibinfo {author} {\bibfnamefont {M.}~\bibnamefont
  {Weides}}, \bibinfo {author} {\bibfnamefont {J.}~\bibnamefont {Wenner}},
  \bibinfo {author} {\bibfnamefont {T.}~\bibnamefont {Yamamoto}}, \bibinfo
  {author} {\bibfnamefont {Y.}~\bibnamefont {Yin}}, \bibinfo {author}
  {\bibfnamefont {J.}~\bibnamefont {Zhao}}, \bibinfo {author} {\bibfnamefont
  {A.~N.}\ \bibnamefont {Cleland}}, \ and\ \bibinfo {author} {\bibfnamefont
  {J.~M.}\ \bibnamefont {Martinis}},\ }\href {\doibase
  10.1103/PhysRevB.84.024501} {\bibfield  {journal} {\bibinfo  {journal} {Phys.
  Rev. B}\ }\textbf {\bibinfo {volume} {84}},\ \bibinfo {pages} {024501}
  (\bibinfo {year} {2011})}\BibitemShut {NoStop}%
\bibitem [{\citenamefont {Wenner}\ \emph {et~al.}(2013)\citenamefont {Wenner},
  \citenamefont {Yin}, \citenamefont {Lucero}, \citenamefont {Barends},
  \citenamefont {Chen}, \citenamefont {Chiaro}, \citenamefont {Kelly},
  \citenamefont {Lenander}, \citenamefont {Mariantoni}, \citenamefont
  {Megrant}, \citenamefont {Neill}, \citenamefont {O'Malley}, \citenamefont
  {Sank}, \citenamefont {Vainsencher}, \citenamefont {Wang}, \citenamefont
  {White}, \citenamefont {Cleland},\ and\ \citenamefont
  {Martinis}}]{Wenner2013}%
  \BibitemOpen
  \bibfield  {author} {\bibinfo {author} {\bibfnamefont {J.}~\bibnamefont
  {Wenner}}, \bibinfo {author} {\bibfnamefont {Y.}~\bibnamefont {Yin}},
  \bibinfo {author} {\bibfnamefont {E.}~\bibnamefont {Lucero}}, \bibinfo
  {author} {\bibfnamefont {R.}~\bibnamefont {Barends}}, \bibinfo {author}
  {\bibfnamefont {Y.}~\bibnamefont {Chen}}, \bibinfo {author} {\bibfnamefont
  {B.}~\bibnamefont {Chiaro}}, \bibinfo {author} {\bibfnamefont
  {J.}~\bibnamefont {Kelly}}, \bibinfo {author} {\bibfnamefont
  {M.}~\bibnamefont {Lenander}}, \bibinfo {author} {\bibfnamefont
  {M.}~\bibnamefont {Mariantoni}}, \bibinfo {author} {\bibfnamefont
  {A.}~\bibnamefont {Megrant}}, \bibinfo {author} {\bibfnamefont
  {C.}~\bibnamefont {Neill}}, \bibinfo {author} {\bibfnamefont {P.~J.~J.}\
  \bibnamefont {O'Malley}}, \bibinfo {author} {\bibfnamefont {D.}~\bibnamefont
  {Sank}}, \bibinfo {author} {\bibfnamefont {A.}~\bibnamefont {Vainsencher}},
  \bibinfo {author} {\bibfnamefont {H.}~\bibnamefont {Wang}}, \bibinfo {author}
  {\bibfnamefont {T.~C.}\ \bibnamefont {White}}, \bibinfo {author}
  {\bibfnamefont {A.~N.}\ \bibnamefont {Cleland}}, \ and\ \bibinfo {author}
  {\bibfnamefont {J.~M.}\ \bibnamefont {Martinis}},\ }\href {\doibase
  10.1103/PhysRevLett.110.150502} {\bibfield  {journal} {\bibinfo  {journal}
  {Phys. Rev. Lett.}\ }\textbf {\bibinfo {volume} {110}},\ \bibinfo {pages}
  {150502} (\bibinfo {year} {2013})}\BibitemShut {NoStop}%
\bibitem [{\citenamefont {Patel}\ \emph {et~al.}(2017)\citenamefont {Patel},
  \citenamefont {Pechenezhskiy}, \citenamefont {Plourde}, \citenamefont
  {Vavilov},\ and\ \citenamefont {McDermott}}]{Patel2017}%
  \BibitemOpen
  \bibfield  {author} {\bibinfo {author} {\bibfnamefont {U.}~\bibnamefont
  {Patel}}, \bibinfo {author} {\bibfnamefont {I.~V.}\ \bibnamefont
  {Pechenezhskiy}}, \bibinfo {author} {\bibfnamefont {B.~L.~T.}\ \bibnamefont
  {Plourde}}, \bibinfo {author} {\bibfnamefont {M.~G.}\ \bibnamefont
  {Vavilov}}, \ and\ \bibinfo {author} {\bibfnamefont {R.}~\bibnamefont
  {McDermott}},\ }\href {\doibase 10.1103/PhysRevB.96.220501} {\bibfield
  {journal} {\bibinfo  {journal} {Phys. Rev. B}\ }\textbf {\bibinfo {volume}
  {96}},\ \bibinfo {pages} {220501} (\bibinfo {year} {2017})}\BibitemShut
  {NoStop}%
\bibitem [{\citenamefont {Riwar}\ and\ \citenamefont
  {Catelani}(2019)}]{Riwar2019}%
  \BibitemOpen
  \bibfield  {author} {\bibinfo {author} {\bibfnamefont {R.-P.}\ \bibnamefont
  {Riwar}}\ and\ \bibinfo {author} {\bibfnamefont {G.}~\bibnamefont
  {Catelani}},\ }\href {\doibase 10.1103/PhysRevB.100.144514} {\bibfield
  {journal} {\bibinfo  {journal} {Phys. Rev. B}\ }\textbf {\bibinfo {volume}
  {100}},\ \bibinfo {pages} {144514} (\bibinfo {year} {2019})}\BibitemShut
  {NoStop}%
\bibitem [{\citenamefont {Arute}\ \emph {et~al.}(2019)\citenamefont {Arute},
  \citenamefont {Arya}, \citenamefont {Babbush}, \citenamefont {Bacon},
  \citenamefont {Bardin}, \citenamefont {Barends}, \citenamefont {Biswas},
  \citenamefont {Boixo}, \citenamefont {Brandao}, \citenamefont {Buell},
  \citenamefont {Burkett}, \citenamefont {Chen}, \citenamefont {Chen},
  \citenamefont {Chiaro}, \citenamefont {Collins}, \citenamefont {Courtney},
  \citenamefont {Dunsworth}, \citenamefont {Farhi}, \citenamefont {Foxen},
  \citenamefont {Fowler}, \citenamefont {Gidney}, \citenamefont {Giustina},
  \citenamefont {Graff}, \citenamefont {Guerin}, \citenamefont {Habegger},
  \citenamefont {Harrigan}, \citenamefont {Hartmann}, \citenamefont {Ho},
  \citenamefont {Hoffmann}, \citenamefont {Huang}, \citenamefont {Humble},
  \citenamefont {Isakov}, \citenamefont {Jeffrey}, \citenamefont {Jiang},
  \citenamefont {Kafri}, \citenamefont {Kechedzhi}, \citenamefont {Kelly},
  \citenamefont {Klimov}, \citenamefont {Knysh}, \citenamefont {Korotkov},
  \citenamefont {Kostritsa}, \citenamefont {Landhuis}, \citenamefont
  {Lindmark}, \citenamefont {Lucero}, \citenamefont {Lyakh}, \citenamefont
  {Mandr{\`a}}, \citenamefont {McClean}, \citenamefont {McEwen}, \citenamefont
  {Megrant}, \citenamefont {Mi}, \citenamefont {Michielsen}, \citenamefont
  {Mohseni}, \citenamefont {Mutus}, \citenamefont {Naaman}, \citenamefont
  {Neeley}, \citenamefont {Neill}, \citenamefont {Niu}, \citenamefont {Ostby},
  \citenamefont {Petukhov}, \citenamefont {Platt}, \citenamefont {Quintana},
  \citenamefont {Rieffel}, \citenamefont {Roushan}, \citenamefont {Rubin},
  \citenamefont {Sank}, \citenamefont {Satzinger}, \citenamefont {Smelyanskiy},
  \citenamefont {Sung}, \citenamefont {Trevithick}, \citenamefont
  {Vainsencher}, \citenamefont {Villalonga}, \citenamefont {White},
  \citenamefont {Yao}, \citenamefont {Yeh}, \citenamefont {Zalcman},
  \citenamefont {Neven},\ and\ \citenamefont {Martinis}}]{Arute2019}%
  \BibitemOpen
  \bibfield  {author} {\bibinfo {author} {\bibfnamefont {F.}~\bibnamefont
  {Arute}}, \bibinfo {author} {\bibfnamefont {K.}~\bibnamefont {Arya}},
  \bibinfo {author} {\bibfnamefont {R.}~\bibnamefont {Babbush}}, \bibinfo
  {author} {\bibfnamefont {D.}~\bibnamefont {Bacon}}, \bibinfo {author}
  {\bibfnamefont {J.~C.}\ \bibnamefont {Bardin}}, \bibinfo {author}
  {\bibfnamefont {R.}~\bibnamefont {Barends}}, \bibinfo {author} {\bibfnamefont
  {R.}~\bibnamefont {Biswas}}, \bibinfo {author} {\bibfnamefont
  {S.}~\bibnamefont {Boixo}}, \bibinfo {author} {\bibfnamefont {F.~G. S.~L.}\
  \bibnamefont {Brandao}}, \bibinfo {author} {\bibfnamefont {D.~A.}\
  \bibnamefont {Buell}}, \bibinfo {author} {\bibfnamefont {B.}~\bibnamefont
  {Burkett}}, \bibinfo {author} {\bibfnamefont {Y.}~\bibnamefont {Chen}},
  \bibinfo {author} {\bibfnamefont {Z.}~\bibnamefont {Chen}}, \bibinfo {author}
  {\bibfnamefont {B.}~\bibnamefont {Chiaro}}, \bibinfo {author} {\bibfnamefont
  {R.}~\bibnamefont {Collins}}, \bibinfo {author} {\bibfnamefont
  {W.}~\bibnamefont {Courtney}}, \bibinfo {author} {\bibfnamefont
  {A.}~\bibnamefont {Dunsworth}}, \bibinfo {author} {\bibfnamefont
  {E.}~\bibnamefont {Farhi}}, \bibinfo {author} {\bibfnamefont
  {B.}~\bibnamefont {Foxen}}, \bibinfo {author} {\bibfnamefont
  {A.}~\bibnamefont {Fowler}}, \bibinfo {author} {\bibfnamefont
  {C.}~\bibnamefont {Gidney}}, \bibinfo {author} {\bibfnamefont
  {M.}~\bibnamefont {Giustina}}, \bibinfo {author} {\bibfnamefont
  {R.}~\bibnamefont {Graff}}, \bibinfo {author} {\bibfnamefont
  {K.}~\bibnamefont {Guerin}}, \bibinfo {author} {\bibfnamefont
  {S.}~\bibnamefont {Habegger}}, \bibinfo {author} {\bibfnamefont {M.~P.}\
  \bibnamefont {Harrigan}}, \bibinfo {author} {\bibfnamefont {M.~J.}\
  \bibnamefont {Hartmann}}, \bibinfo {author} {\bibfnamefont {A.}~\bibnamefont
  {Ho}}, \bibinfo {author} {\bibfnamefont {M.}~\bibnamefont {Hoffmann}},
  \bibinfo {author} {\bibfnamefont {T.}~\bibnamefont {Huang}}, \bibinfo
  {author} {\bibfnamefont {T.~S.}\ \bibnamefont {Humble}}, \bibinfo {author}
  {\bibfnamefont {S.~V.}\ \bibnamefont {Isakov}}, \bibinfo {author}
  {\bibfnamefont {E.}~\bibnamefont {Jeffrey}}, \bibinfo {author} {\bibfnamefont
  {Z.}~\bibnamefont {Jiang}}, \bibinfo {author} {\bibfnamefont
  {D.}~\bibnamefont {Kafri}}, \bibinfo {author} {\bibfnamefont
  {K.}~\bibnamefont {Kechedzhi}}, \bibinfo {author} {\bibfnamefont
  {J.}~\bibnamefont {Kelly}}, \bibinfo {author} {\bibfnamefont {P.~V.}\
  \bibnamefont {Klimov}}, \bibinfo {author} {\bibfnamefont {S.}~\bibnamefont
  {Knysh}}, \bibinfo {author} {\bibfnamefont {A.}~\bibnamefont {Korotkov}},
  \bibinfo {author} {\bibfnamefont {F.}~\bibnamefont {Kostritsa}}, \bibinfo
  {author} {\bibfnamefont {D.}~\bibnamefont {Landhuis}}, \bibinfo {author}
  {\bibfnamefont {M.}~\bibnamefont {Lindmark}}, \bibinfo {author}
  {\bibfnamefont {E.}~\bibnamefont {Lucero}}, \bibinfo {author} {\bibfnamefont
  {D.}~\bibnamefont {Lyakh}}, \bibinfo {author} {\bibfnamefont
  {S.}~\bibnamefont {Mandr{\`a}}}, \bibinfo {author} {\bibfnamefont {J.~R.}\
  \bibnamefont {McClean}}, \bibinfo {author} {\bibfnamefont {M.}~\bibnamefont
  {McEwen}}, \bibinfo {author} {\bibfnamefont {A.}~\bibnamefont {Megrant}},
  \bibinfo {author} {\bibfnamefont {X.}~\bibnamefont {Mi}}, \bibinfo {author}
  {\bibfnamefont {K.}~\bibnamefont {Michielsen}}, \bibinfo {author}
  {\bibfnamefont {M.}~\bibnamefont {Mohseni}}, \bibinfo {author} {\bibfnamefont
  {J.}~\bibnamefont {Mutus}}, \bibinfo {author} {\bibfnamefont
  {O.}~\bibnamefont {Naaman}}, \bibinfo {author} {\bibfnamefont
  {M.}~\bibnamefont {Neeley}}, \bibinfo {author} {\bibfnamefont
  {C.}~\bibnamefont {Neill}}, \bibinfo {author} {\bibfnamefont {M.~Y.}\
  \bibnamefont {Niu}}, \bibinfo {author} {\bibfnamefont {E.}~\bibnamefont
  {Ostby}}, \bibinfo {author} {\bibfnamefont {A.}~\bibnamefont {Petukhov}},
  \bibinfo {author} {\bibfnamefont {J.~C.}\ \bibnamefont {Platt}}, \bibinfo
  {author} {\bibfnamefont {C.}~\bibnamefont {Quintana}}, \bibinfo {author}
  {\bibfnamefont {E.~G.}\ \bibnamefont {Rieffel}}, \bibinfo {author}
  {\bibfnamefont {P.}~\bibnamefont {Roushan}}, \bibinfo {author} {\bibfnamefont
  {N.~C.}\ \bibnamefont {Rubin}}, \bibinfo {author} {\bibfnamefont
  {D.}~\bibnamefont {Sank}}, \bibinfo {author} {\bibfnamefont {K.~J.}\
  \bibnamefont {Satzinger}}, \bibinfo {author} {\bibfnamefont {V.}~\bibnamefont
  {Smelyanskiy}}, \bibinfo {author} {\bibfnamefont {K.~J.}\ \bibnamefont
  {Sung}}, \bibinfo {author} {\bibfnamefont {M.~D.}\ \bibnamefont
  {Trevithick}}, \bibinfo {author} {\bibfnamefont {A.}~\bibnamefont
  {Vainsencher}}, \bibinfo {author} {\bibfnamefont {B.}~\bibnamefont
  {Villalonga}}, \bibinfo {author} {\bibfnamefont {T.}~\bibnamefont {White}},
  \bibinfo {author} {\bibfnamefont {Z.~J.}\ \bibnamefont {Yao}}, \bibinfo
  {author} {\bibfnamefont {P.}~\bibnamefont {Yeh}}, \bibinfo {author}
  {\bibfnamefont {A.}~\bibnamefont {Zalcman}}, \bibinfo {author} {\bibfnamefont
  {H.}~\bibnamefont {Neven}}, \ and\ \bibinfo {author} {\bibfnamefont {J.~M.}\
  \bibnamefont {Martinis}},\ }\href {\doibase 10.1038/s41586-019-1666-5}
  {\bibfield  {journal} {\bibinfo  {journal} {Nature}\ }\textbf {\bibinfo
  {volume} {574}},\ \bibinfo {pages} {505} (\bibinfo {year}
  {2019})}\BibitemShut {NoStop}%
\bibitem [{\citenamefont {Herr}\ \emph {et~al.}(2007)\citenamefont {Herr},
  \citenamefont {Fedorov}, \citenamefont {Shnirman}, \citenamefont {Il'ichev},\
  and\ \citenamefont {Schön}}]{Herr2007}%
  \BibitemOpen
  \bibfield  {author} {\bibinfo {author} {\bibfnamefont {A.}~\bibnamefont
  {Herr}}, \bibinfo {author} {\bibfnamefont {A.}~\bibnamefont {Fedorov}},
  \bibinfo {author} {\bibfnamefont {A.}~\bibnamefont {Shnirman}}, \bibinfo
  {author} {\bibfnamefont {E.}~\bibnamefont {Il'ichev}}, \ and\ \bibinfo
  {author} {\bibfnamefont {G.}~\bibnamefont {Schön}},\ }\href {\doibase
  10.1088/0953-2048/20/11/s29} {\bibfield  {journal} {\bibinfo  {journal}
  {Supercond. Sci. Tech.}\ }\textbf {\bibinfo {volume} {20}},\ \bibinfo {pages}
  {S450} (\bibinfo {year} {2007})}\BibitemShut {NoStop}%
\bibitem [{\citenamefont {Fedorov}\ \emph {et~al.}(2014)\citenamefont
  {Fedorov}, \citenamefont {Shcherbakova}, \citenamefont {Wolf}, \citenamefont
  {Beckmann},\ and\ \citenamefont {Ustinov}}]{Fedorov2014}%
  \BibitemOpen
  \bibfield  {author} {\bibinfo {author} {\bibfnamefont {K.~G.}\ \bibnamefont
  {Fedorov}}, \bibinfo {author} {\bibfnamefont {A.~V.}\ \bibnamefont
  {Shcherbakova}}, \bibinfo {author} {\bibfnamefont {M.~J.}\ \bibnamefont
  {Wolf}}, \bibinfo {author} {\bibfnamefont {D.}~\bibnamefont {Beckmann}}, \
  and\ \bibinfo {author} {\bibfnamefont {A.~V.}\ \bibnamefont {Ustinov}},\
  }\href {\doibase 10.1103/PhysRevLett.112.160502} {\bibfield  {journal}
  {\bibinfo  {journal} {Phys. Rev. Lett.}\ }\textbf {\bibinfo {volume} {112}},\
  \bibinfo {pages} {160502} (\bibinfo {year} {2014})}\BibitemShut {NoStop}%
\bibitem [{\citenamefont {Howington}\ \emph {et~al.}(2019)\citenamefont
  {Howington}, \citenamefont {Opremcak}, \citenamefont {McDermott},
  \citenamefont {Kirichenko}, \citenamefont {Mukhanov},\ and\ \citenamefont
  {Plourde}}]{Howington2019}%
  \BibitemOpen
  \bibfield  {author} {\bibinfo {author} {\bibfnamefont {C.}~\bibnamefont
  {Howington}}, \bibinfo {author} {\bibfnamefont {A.}~\bibnamefont {Opremcak}},
  \bibinfo {author} {\bibfnamefont {R.}~\bibnamefont {McDermott}}, \bibinfo
  {author} {\bibfnamefont {A.}~\bibnamefont {Kirichenko}}, \bibinfo {author}
  {\bibfnamefont {O.~A.}\ \bibnamefont {Mukhanov}}, \ and\ \bibinfo {author}
  {\bibfnamefont {B.~L.~T.}\ \bibnamefont {Plourde}},\ }\href {\doibase
  10.1109/tasc.2019.2908884} {\bibfield  {journal} {\bibinfo  {journal} {IEEE
  Trans. Appl. Supercond.}\ }\textbf {\bibinfo {volume} {29}},\ \bibinfo
  {pages} {1} (\bibinfo {year} {2019})}\BibitemShut {NoStop}%
\bibitem [{\citenamefont {McDermott}\ and\ \citenamefont
  {Vavilov}(2014)}]{McDermott2014}%
  \BibitemOpen
  \bibfield  {author} {\bibinfo {author} {\bibfnamefont {R.}~\bibnamefont
  {McDermott}}\ and\ \bibinfo {author} {\bibfnamefont {M.~G.}\ \bibnamefont
  {Vavilov}},\ }\href {\doibase 10.1103/PhysRevApplied.2.014007} {\bibfield
  {journal} {\bibinfo  {journal} {Phys. Rev. Applied}\ }\textbf {\bibinfo
  {volume} {2}},\ \bibinfo {pages} {014007} (\bibinfo {year}
  {2014})}\BibitemShut {NoStop}%
\bibitem [{\citenamefont {Leonard}\ \emph {et~al.}(2019)\citenamefont
  {Leonard}, \citenamefont {Beck}, \citenamefont {Nelson}, \citenamefont
  {Christensen}, \citenamefont {Thorbeck}, \citenamefont {Howington},
  \citenamefont {Opremcak}, \citenamefont {Pechenezhskiy}, \citenamefont
  {Dodge}, \citenamefont {Dupuis}, \citenamefont {Hutchings}, \citenamefont
  {Ku}, \citenamefont {Schlenker}, \citenamefont {Suttle}, \citenamefont
  {Wilen}, \citenamefont {Zhu}, \citenamefont {Vavilov}, \citenamefont
  {Plourde},\ and\ \citenamefont {McDermott}}]{Leonard2019}%
  \BibitemOpen
  \bibfield  {author} {\bibinfo {author} {\bibfnamefont {E.}~\bibnamefont
  {Leonard}}, \bibinfo {author} {\bibfnamefont {M.~A.}\ \bibnamefont {Beck}},
  \bibinfo {author} {\bibfnamefont {J.}~\bibnamefont {Nelson}}, \bibinfo
  {author} {\bibfnamefont {B.~G.}\ \bibnamefont {Christensen}}, \bibinfo
  {author} {\bibfnamefont {T.}~\bibnamefont {Thorbeck}}, \bibinfo {author}
  {\bibfnamefont {C.}~\bibnamefont {Howington}}, \bibinfo {author}
  {\bibfnamefont {A.}~\bibnamefont {Opremcak}}, \bibinfo {author}
  {\bibfnamefont {I.~V.}\ \bibnamefont {Pechenezhskiy}}, \bibinfo {author}
  {\bibfnamefont {K.}~\bibnamefont {Dodge}}, \bibinfo {author} {\bibfnamefont
  {N.~P.}\ \bibnamefont {Dupuis}}, \bibinfo {author} {\bibfnamefont {M.~D.}\
  \bibnamefont {Hutchings}}, \bibinfo {author} {\bibfnamefont {J.}~\bibnamefont
  {Ku}}, \bibinfo {author} {\bibfnamefont {F.}~\bibnamefont {Schlenker}},
  \bibinfo {author} {\bibfnamefont {J.}~\bibnamefont {Suttle}}, \bibinfo
  {author} {\bibfnamefont {C.}~\bibnamefont {Wilen}}, \bibinfo {author}
  {\bibfnamefont {S.}~\bibnamefont {Zhu}}, \bibinfo {author} {\bibfnamefont
  {M.}~\bibnamefont {Vavilov}}, \bibinfo {author} {\bibfnamefont {B.~L.~T.}\
  \bibnamefont {Plourde}}, \ and\ \bibinfo {author} {\bibfnamefont
  {R.}~\bibnamefont {McDermott}},\ }\href {\doibase
  10.1103/PhysRevApplied.11.014009} {\bibfield  {journal} {\bibinfo  {journal}
  {Phys. Rev. Applied}\ }\textbf {\bibinfo {volume} {11}},\ \bibinfo {pages}
  {014009} (\bibinfo {year} {2019})}\BibitemShut {NoStop}%
\bibitem [{\citenamefont {Dolan}(1977)}]{Dolan1977}%
  \BibitemOpen
  \bibfield  {author} {\bibinfo {author} {\bibfnamefont {G.~J.}\ \bibnamefont
  {Dolan}},\ }\href {\doibase 10.1063/1.89690} {\bibfield  {journal} {\bibinfo
  {journal} {Appl. Phys. Lett.}\ }\textbf {\bibinfo {volume} {31}},\ \bibinfo
  {pages} {337} (\bibinfo {year} {1977})}\BibitemShut {NoStop}%
\bibitem [{\citenamefont {Schuster}\ \emph {et~al.}(2005)\citenamefont
  {Schuster}, \citenamefont {Wallraff}, \citenamefont {Blais}, \citenamefont
  {Frunzio}, \citenamefont {Huang}, \citenamefont {Majer}, \citenamefont
  {Girvin},\ and\ \citenamefont {Schoelkopf}}]{Schuster2005}%
  \BibitemOpen
  \bibfield  {author} {\bibinfo {author} {\bibfnamefont {D.~I.}\ \bibnamefont
  {Schuster}}, \bibinfo {author} {\bibfnamefont {A.}~\bibnamefont {Wallraff}},
  \bibinfo {author} {\bibfnamefont {A.}~\bibnamefont {Blais}}, \bibinfo
  {author} {\bibfnamefont {L.}~\bibnamefont {Frunzio}}, \bibinfo {author}
  {\bibfnamefont {R.-S.}\ \bibnamefont {Huang}}, \bibinfo {author}
  {\bibfnamefont {J.}~\bibnamefont {Majer}}, \bibinfo {author} {\bibfnamefont
  {S.~M.}\ \bibnamefont {Girvin}}, \ and\ \bibinfo {author} {\bibfnamefont
  {R.~J.}\ \bibnamefont {Schoelkopf}},\ }\href {\doibase
  10.1103/PhysRevLett.94.123602} {\bibfield  {journal} {\bibinfo  {journal}
  {Phys. Rev. Lett.}\ }\textbf {\bibinfo {volume} {94}},\ \bibinfo {pages}
  {123602} (\bibinfo {year} {2005})}\BibitemShut {NoStop}%
\bibitem [{\citenamefont {Schuster}\ \emph {et~al.}(2007)\citenamefont
  {Schuster}, \citenamefont {Houck}, \citenamefont {Schreier}, \citenamefont
  {Wallraff}, \citenamefont {Gambetta}, \citenamefont {Blais}, \citenamefont
  {Frunzio}, \citenamefont {Majer}, \citenamefont {Johnson}, \citenamefont
  {Devoret}, \citenamefont {Girvin},\ and\ \citenamefont
  {Schoelkopf}}]{Schuster2007}%
  \BibitemOpen
  \bibfield  {author} {\bibinfo {author} {\bibfnamefont {D.~I.}\ \bibnamefont
  {Schuster}}, \bibinfo {author} {\bibfnamefont {A.~A.}\ \bibnamefont {Houck}},
  \bibinfo {author} {\bibfnamefont {J.~A.}\ \bibnamefont {Schreier}}, \bibinfo
  {author} {\bibfnamefont {A.}~\bibnamefont {Wallraff}}, \bibinfo {author}
  {\bibfnamefont {J.~M.}\ \bibnamefont {Gambetta}}, \bibinfo {author}
  {\bibfnamefont {A.}~\bibnamefont {Blais}}, \bibinfo {author} {\bibfnamefont
  {L.}~\bibnamefont {Frunzio}}, \bibinfo {author} {\bibfnamefont
  {J.}~\bibnamefont {Majer}}, \bibinfo {author} {\bibfnamefont
  {B.}~\bibnamefont {Johnson}}, \bibinfo {author} {\bibfnamefont {M.~H.}\
  \bibnamefont {Devoret}}, \bibinfo {author} {\bibfnamefont {S.~M.}\
  \bibnamefont {Girvin}}, \ and\ \bibinfo {author} {\bibfnamefont {R.~J.}\
  \bibnamefont {Schoelkopf}},\ }\href {\doibase 10.1038/nature05461} {\bibfield
   {journal} {\bibinfo  {journal} {Nature}\ }\textbf {\bibinfo {volume}
  {445}},\ \bibinfo {pages} {515} (\bibinfo {year} {2007})}\BibitemShut
  {NoStop}%
\end{thebibliography}%

\end{document}